\pdfoutput=1
\documentclass[11pt]{article}
\usepackage{graphicx}
\usepackage{latexsym}
\usepackage{mathrsfs}
\usepackage[cal=boondox]{mathalfa}
\usepackage[overload]{textcase}

.5 scaled \magstep4
\font\medio=cmr9.5 scaled \magstep2
\outer\def\beginsection#1\par{\medbreak\bigskip
      \message{#1}\leftline{\bf#1}\nobreak\medskip
\vskip-\parskip
      \noindent}
\begin{document}
\bibliographystyle{unsrt}

\titlepage
\vspace{1cm}
\begin{center}
{\Large {\bf Ultra-high frequency spikes of relic gravitons}}\\
\vspace{1.5 cm}
Massimo Giovannini \footnote{e-mail address: massimo.giovannini@cern.ch}\\
\vspace{1cm}
{{\sl Department of Physics, CERN, 1211 Geneva 23, Switzerland }}\\
\vspace{0.5cm}
{{\sl INFN, Section of Milan-Bicocca, 20126 Milan, Italy}}
\vspace*{1cm}
\end{center}
\vskip 0.3cm
\centerline{\medio  Abstract}
\vskip 0.5cm
The maximal frequency domain of the cosmic gravitons falls in the THz region where, without conflicting with the existing phenomenological bounds, only few particles  with opposite (comoving) three-momenta are produced.
Although any reliable scrutiny of the ultra-high frequency spikes should include all the sources of late-time suppression at lower and intermediate frequencies, some relevant properties of the averaged multiplicities and of the spectral energy density can be derived within a reduced set of approximations that may become invalid as the frequency decreases well below the Hz. The accuracy of these concurrent approaches is assessed from the properties of the transition matrix that relates the late-time spectra to the values of the mode functions during an inflationary stage. In the obtained framework the bounds on the post-inflationary expansion rate are swiftly deduced and compare quite well with the ones including a more faithful numerical treatment. It also follows that the timeline of the post-inflationary expansion rate might be observationally accessible, in the years to come, provided the electromechanical detectors (like microwave cavities or waveguides) operating between the MHz and the THz shall eventually reach sensitivities in the chirp amplitudes which are (at least) twelve orders of magnitude smaller than the ones experimentally attainable in the audio band (i.e. between few Hz and the kHz). 
\noindent
\vspace{5mm}
\vfill
\newpage
\renewcommand{\theequation}{1.\arabic{equation}}
\setcounter{equation}{0}
\section{Introduction}
\label{sec1}
The low frequency gravitational waves constrain the conventional 
inflationary scenarios through the relative weight of the tensor and scalar power spectra (denoted hereunder by $r_{T}$) at the conventional pivot scale $k_{p}$ \cite{SW1}. According to the available observational limits we should require that $r_{T} < 0.06$ \cite{TT1,TT2,TT3} and since the frequency corresponding to $k_{p}= 0.002\,\, \mathrm{Mpc}^{-1}$ is of the order of $\nu_{p} = 3.09\, \mathrm{aHz}$ (where $1\, \mathrm{aHz} = 10^{-18}$ Hz) the bounds on $r_{T}$ ultimately constrain the low frequency spectral range of the relic gravitons (see, for instance, \cite{MG1} for a review). Indeed, even before the formulation of the conventional inflationary paradigm it has been realized that any variation of the space-time curvature must produce shots of relic gravitons \cite{AA1,AA1a} with different averaged multiplicities and slopes that faithfully reflect the early evolution of the background geometry. A standard timeline of the expansion rate would stipulate that an inflationary stage is followed by a radiation-dominated epoch so that, at the present conformal time $\tau_{0}$, the spectral energy density in critical units  (denoted in what follows by $\Omega_{gw}(\nu,\tau_{0})$) is quasi-flat for comoving frequencies $\nu$ approximately larger than $100$ aHz \cite{AA2,AA3}; more specifically in this spectral domain we should have that $h_{0}^2 \Omega_{gw}(\nu,\tau_{0}) < {\mathcal O}(10^{-17})$ (where $h_{0}$ denotes, in what follows, the indetermination in the current value of the Hubble rate). In the concordance paradigm (possibly supplemented by an early stage of inflationary expansion) the  spectral slope always decreases and thanks to the consistency relations, the tensor spectral index $n_{T}$ is notoriously related to the tensor to scalar ratio $r_{T}$ as $n_{T} \simeq - r_{T}/8$ \cite{SW1,TT1,TT2,TT3}. This conclusion can be however evaded provided the post-inflationary expansion rate is slower than radiation as originally suggested in Ref. \cite{EXP1}: in this case $\Omega_{gw}(\nu,\tau_{0})$ may generically increase as a function of the comoving frequency. Depending on the timeline of the expansion rate, the presence of spikes may arise in various ranges either below the Hz or above it.  Although the considerations discussed here may be applied to a number of physical situations, our examples primarily involve the presence of a single post-inflationary phase different from the radiation stage that may be recovered 
at a later (or even much later) epoch.

If the spectral energy density increases either in the intermediate or high frequency ranges, 
the expected signals in the MHz and GHz domains may even be eight orders of magnitude larger than the ones of the concordance paradigm \cite{MG1,EXP1} and this is why electromechanical  detectors are believed to be particularly suitable for the detection of high frequency gravitons. While the first detectors proposed at high frequencies consisted of toroidal waveguides with a propagating wavepacket of electromagnetic radiation \cite{BB1,BB2}, also static electromagnetic fields (i.e. microwave cavities with superconducting walls) have been considered later on \cite{CAV1,CAV2,CAV3,CAV4,CAV5,CAV6}. The original  prototypes could in principle resolve chirp amplitudes $h_{c} ={\mathcal O}(10^{-17})$ \cite{CAV1,CAV2,CAV3} and nearly twenty years later the sensitivities reached the level of $h_{c} = {\mathcal O}(10^{-20})$  \cite{CAV4,CAV5}. The minimal chirp amplitudes resolved by these detectors might be today  comparable with the ones probed by wide-band interferometers in a much lower frequency range. In principle microwave cavities could detect relic gravitons between few GHz and $0.1$ THz  \cite{CAV6} but other devices have been suggested in similar ranges \cite{CAV7,CAV8,CAV9}. 
  
While the lowest frequency of the relic gravitons is ultimately determined by 
the current value of the Hubble rate, the maximal frequency apparently depends non-trivially on the specific scenario \cite{MAX1}. For instance in the concordance paradigm the maximal frequency turns out to be ${\mathcal O}(100)$ MHz but this observation is purely academic: because of the steady decrease of the spectral energy density,  the largest signal occurs anyway close to the pivot frequency $\nu_{p} = {\mathcal O}(\mathrm{aHz})$. Absent any direct or indirect phenomenological indication on the post-inflationary history prior to the onset of big bang nucleosynthesis, the early timeline of the expansion rate may well be a sequel of arbitrary stages fulfilling certain reasonable constraints \cite{MAX2}.  Broadly speaking, every phase expanding faster than radiation reduces the value of the maximal frequency below ${\mathcal O}(100)$ MHz while the opposite is true for periods that expand less rapidly than radiation. If the expansion is slower than radiation for a sizeable portion of the post-inflationary evolution the maximal frequency may even exceed the GHz \cite{MAX1,MAX2}. The indetermination of the maximal frequency is related to the model dependence of the minimal wavelength below which the ultraviolet modes are never stretched by the inflationary expansion. Although the actual value of the maximal frequency depends upon the timeline of the expansion rate, it is however possible to deduce a general bound that is independent on the specific scenario \cite{MAX1}. For this purpose we may observe that the spectral energy density in critical units ultimately depends on $\overline{n}(\nu, \tau_{0})$ that denotes throughout the averaged multiplicity of the produced gravitons as a function of the (comoving) frequency: 
\begin{equation}
\Omega_{gw}(\nu, \tau_{0}) = \frac{128\, \pi^3}{3} \,\, \frac{\nu^{4}}{H_{0}^2 \, M_{P}^2}\,\, \overline{n}(\nu, \tau_{0}),
\label{FF1}
\end{equation}
where $H_{0}$ is the current value of the Hubble rate and $M_{P} = 1/\sqrt{G}$ is the Planck mass\footnote{In this paper $M_{P} = 1/\sqrt{G}$  while $\overline{M}_{P} = 1/\sqrt{8\pi G}$ is the reduced Planck mass. Within the present notations we also introduce the Planck length $\ell_{P} = 1/\overline{M}_{P}$.} in the units adopted here (i.e. $\hbar=c=1$);  $\tau_{0}$ denotes throughout the present value of the conformal time coordinate. Because of the unitarity of the process of graviton production, for $\nu > \nu_{max}$ the averaged multiplicity gets exponentially suppressed \cite{AA01,AA02} (see also \cite{BIRREL,PTOMS,FORDH,FORDL}) at a rate that depends on the smoothness of the transition. For $\nu > \nu_{max}$ the averaged multiplicity scales as $e^{- \gamma (\nu/\nu_{max})}$ where the value of $\gamma$ is ${\mathcal O}(1)$ and can be concretely determined, for the specific scenario under consideration, with direct numerical integration as discussed in the past for the stochastic backgrounds of relic gravitational radiation \cite{mg0}. Around the maximal frequency domain only few graviton pairs are produced and this means that in the limit $\nu\to \nu_{max}$ the averaged multiplicity $\overline{n}(\nu_{max}, \tau_{0}) = {\mathcal O}(1)$; from Eq. (\ref{FF1}) the high frequency limits applicable to $\Omega_{gw}(\nu_{max}, \tau_{0})$ can be translated into bounds on $\nu_{max}$ with the result that \cite{MAX1}
\begin{equation}
\nu_{max} < {\mathcal O}(0.1)\,\, \Omega_{R0}^{1/4}\,\, \sqrt{H_{0}\, M_{P}} < \,\mathrm{THz},
\label{FF2}
\end{equation}
where $\Omega_{R0}$ denotes throughout the current fraction of relativistic species in the plasma. In the minimal version of the concordance paradigm the neutrinos are often considered to be massless since they practically count as radiation when the initial conditions 
of CMB anisotropies are set; thus $h_{0}^2 \Omega_{R0}$ ranges in practice between $2.47\times 10^{-5}$ and $4.15\times 10^{-5}$ \cite{SW1} (see also \cite{TT1,TT2,TT3}). The difference between these figures is immaterial for the present ends but it should be borne in mind.

The careful evaluation of the spectrum of the relic gravitons must also include various sources of late-time suppression and in the concordance paradigm the spectral energy density is reduced by the transition to the dark energy, by the evolution of the relativistic species  and, most notably, by the free-streaming of neutrinos \cite{STRNU0,STRNU1,STRNU2,STRNU3,STRNU4}. The free-streaming is effective below the nHz,  i.e. close to the typical frequency associated with big bang nucleosynthesis (see also \cite{MG1} and discussions therein). While a rigorous and quantitative perspective is unavoidable (and has been thoroughly pursued in the relatively recent past) we intend to introduce here an approximate (but semi-analytic) discussion of the high frequency spike of the spectral energy density for two complementary reasons. In the first place simpler (but approximate) expressions are useful for the derivation of specific templates and are also relevant if we ought to set the accuracy of different approximation schemes. The second motivation is that, within one order of magnitude, the high frequency limits are independent on the damping effects occurring at smaller frequencies even though the late-time suppression does affect the overall normalization of the spectrum. This is the logic developed hereunder and the layout of the investigation is, in short, the following. In section \ref{sec2} some of the technical aspects of the problem are outlined and briefly discussed with particular attention to the general form of the observables that are specifically 
analyzed and computed in the different sections of the paper. The evaluation of the high frequency spectrum is explored in section \ref{sec3} within the Wentzel–Kramers–Brillouin (WKB) approach. In section \ref{sec4} the spectra are instead deduced from the elements of a transition matrix relating the late-time values of the mode functions to the inflationary spectra. Even if, a posteriori, the latter strategy turns out to be more accurate than the former, in both cases the analysis involves generic post-inflationary expansion rates that do not necessarily coincide with radiation. Particular attention is paid to the case where the rate is slower than radiation and the underlying sources are maximally stiff. After a quantitative comparison of the different approaches, in section \ref{sec5} the high and low frequency normalizations are compared both analytically and numerically. Once the bound of Eq. (\ref{FF2}) is enforced the two strategies are fully compatible but the analytic discussion is swifter than the numerical one which is however intrinsically more accurate in the realistic situation. At the end of section \ref{sec5} the typical values of the chirp amplitudes are discussed with few comments on the detectability strategies that often disregard the physical properties of the expected signals. Section \ref{sec6} contains our concluding considerations.

\renewcommand{\theequation}{2.\arabic{equation}}
\setcounter{equation}{0}
\section{Random backgrounds of relic gravitons}
\label{sec2}
To avoid extended introductions that can be found elsewhere (see e.g. \cite{MAX1,MAX2}), 
we directly start from the Hamiltonian of the relic gravitons
written in terms of the canonical fields $h_{i\,j}(\vec{x},\tau)$ and of the corresponding canonical momenta $\pi_{i\,j}(\vec{x},\tau)$
\begin{equation}
H_{g}(\tau) =  \int d^{3} x \biggl[ \frac{8 \ell_{P}^2}{a^2(\tau)} \pi_{i\,j}\, \pi^{i\,j} + 
\frac{a^2(\tau)}{8 \ell_{P}^2} \partial_{k} h_{i\, j} \partial^{k} h^{i\,j} \biggr],
\label{ONEA}
\end{equation}
where $\ell_{P} = \sqrt{8 \pi G}$ and $a(\tau)$ is the scale factor of a spatially flat 
Friedmann-Robertson-Walker metric expressed as a function of the conformal time coordinate $\tau$. Both $h_{i\,j}(\vec{x},\tau)$ and $\pi_{i\,j}(\vec{x},\tau)$ are solenoidal and traceless and describe the tensor inhomogeneities of the geometry. Furthermore Eq. (\ref{ONEA}) directly follows from the effective action discussed, with different strategies, in Refs. \cite{AC1,AC2} and since $H_{g}(\tau)$ is explicitly dependent upon the conformal time coordinate, all other Hamiltonians (possibly related 
to Eq. (\ref{ONEA}) via a time-dependent canonical transformation) are equally 
acceptable from the classical viewpoint but may lead to potential ambiguities at the quantum level. From Eq. (\ref{ONEA}) the evolution of $\pi_{i\, j}(\vec{x},\tau)$ and $h_{i\,j}(\vec{x},\tau)$ can be written as: 
\begin{equation}
\partial_{\tau} \, h_{i\,j} = \frac{8 \ell_{P}^2}{a^2(\tau)} \, \pi_{i\, j}, \qquad \partial_{\tau} \pi_{i\,j} = \frac{a^2(\tau)}{8 \, \ell_{P}^2} \nabla^2 h_{i\,j}.
\label{ONEB}
\end{equation}
We shall be generically considering hereunder a timeline of the curvature scale where a conventional inflationary stage precedes a phase of decelerated expansion which does not necessarily coincide with radiation. Therefore provided the inflationary epoch is long enough the classical inhomogeneities are suppressed and while the quantum fluctuations keep on reappearing all the time; $h_{i\,j}(\vec{x},\tau)$ and $\pi_{i\,j}(\vec{x},\tau)$ must then be associated with the appropriate Hermitian field operators:
\begin{eqnarray}
\widehat{h}_{i\,j}(\vec{x},\tau) &=& \frac{\sqrt{2} \, \ell_{P}}{(2\pi)^{3/2}} \sum_{\alpha} \, \int\, d^{3} k\, e_{i\,j}^{(\alpha)}(\hat{k})
\biggl[ \widehat{b}_{\vec{k},\, \alpha} \, F_{k,\, \alpha}(\tau) e^{- i \vec{k}\cdot\vec{x}} + \mathrm{H.\,c.}\biggr],
\label{ONEC}\\
\widehat{\pi}_{i\,j}(\vec{x},\tau) &=& \frac{a^2}{4\,\sqrt{2}\, \ell_{P}\, (2\pi)^{3/2}} 
\sum_{\alpha} \, \int\, d^{3} k \, e_{i\,j}^{(\alpha)}(\hat{k}) \biggl[ \widehat{b}_{\vec{k},\, \alpha} \, G_{k,\, \alpha}(\tau) e^{- i \vec{k}\cdot\vec{x}} + \mathrm{H.\,c.}\biggr].
\label{ONED}
\end{eqnarray}
Both in Eqs. (\ref{ONEC}) and (\ref{ONED}) ``H.c.'' indicates the Hermitian conjugate of the preceding term
and the sum over $\alpha$ runs over the two tensor polarizations; if we introduce a triplet of unit vectors 
denoted, respectively, as ($\hat{m}$, $\hat{n}$ and $\hat{k}$ with $\hat{m} \times \hat{n} = \hat{k}$) 
the two tensor polarizations are defined as
$e_{i\,j}^{(\oplus)}(\hat{k}) = (\hat{m}_{i} \hat{m}_{j} - \hat{n}_{i} \hat{n}_{j})$ 
and $e_{i\,j}^{(\otimes)}(\hat{k}) = (\hat{m}_{i} \hat{n}_{j} - \hat{n}_{i} \hat{m}_{j})$. The commutation relations 
between the creation and annihilation operators are defined for a continuum of momenta, i.e. $[\widehat{b}_{\vec{k},\, \alpha}, \widehat{b}_{\vec{p},\, \beta}^{\, \dagger}] = \delta_{\alpha\beta}\, \delta^{(3)}(\vec{k} - \vec{p})$. Finally, the mode functions $F_{k,\,\alpha}$ and $G_{k,\,\alpha}$ introduced in Eqs. (\ref{ONEC})--(\ref{ONED}) can be rescaled at wish; a useful parametrization is  $a\,F_{k,\,\alpha} = f_{k,\,\alpha} $ and $a\,G_{k,\,\alpha} = g_{k,\,\alpha}$. Thus the evolution of  $F_{k,\,\alpha}$ and $G_{k,\,\alpha}$ follows directly from $f_{k,\,\alpha}$ and $g_{k,\,\alpha}$ since the scale factor will always be explicitly continuous with its first derivative; the equations obeyed by 
$f_{k,\,\alpha}$ and $g_{k,\,\alpha}$ follows by inserting Eqs. (\ref{ONEC})--(\ref{ONED}) into Eq. (\ref{ONEB}): 
\begin{equation}
f_{k}^{\prime} = g_{k} + {\mathcal H} \, f_{k}, \qquad g_{k}^{\prime} = - {\mathcal H} g_{k} - k^2 f_{k},\qquad {\mathcal H} = a^{\prime}/a.
\label{TWOA}
\end{equation}
In Eq. (\ref{TWOA}) the prime denotes a derivation with respect to the conformal time coordinate 
$\tau$ and the index $\alpha$ has been suppressed since the same equations 
hold for both polarizations. The field operators of Eqs. (\ref{ONEC})--(\ref{ONED}) can also be represented in Fourier space 
\begin{equation}
\widehat{h}_{i\,j}(\vec{k},\tau) = \frac{1}{(2\pi)^{3/2}} \int d^{3} x \,\,\widehat{h}_{i\,j}(\vec{x},\tau)\, e^{i \vec{k}\cdot\vec{x}}, \qquad \widehat{\pi}_{i\,j}(\vec{k},\tau) = \frac{1}{(2\pi)^{3/2}} \int d^{3} x \,\,\widehat{\pi}_{i\,j}(\vec{x},\tau)\, e^{i \vec{k}\cdot\vec{x}}.
\label{TWOA1}
\end{equation}
From Eqs. (\ref{ONEC})--(\ref{ONED}) and (\ref{TWOA1}) the commutation relations of the field operators at equal times become
\begin{equation}
[\widehat{h}_{i\,j}(\vec{q},\tau), \, \widehat{\pi}_{m\,n}(\vec{p},\tau)] = i \, {\mathcal S}_{i\,j\,
m\,n}(\hat{q}) \, \delta^{(3)}(\vec{q} + \vec{p}),
\label{TWOA2}
\end{equation}
where, as usual, ${\mathcal S}_{i\,j\,m\,n}(\hat{q}) = [p_{i\,m}(\hat{q}) p_{j\,n}(\hat{q}) + 
p_{i\,n}(\hat{q}) p_{j\,n}(\hat{q}) - p_{i\,j}(\hat{q}) p_{m\,n}(\hat{q}) ]/4$ 
and $p_{i\, j}(\hat{q}) = (\delta_{i\,j} - \hat{q}_{i} \, \hat{q}_{j})$ indicates  
the traceless projector. If the canonical form 
of the commutation relations (\ref{TWOA2}) is to be preserved by the time evolution 
of the field operators (\ref{ONEC})--(\ref{ONED}), the mode 
functions are imperatively subjected to the Wronskian normalization condition 
\begin{equation}
f_{k}(\tau)\,g_{k}^{\ast}(\tau) - f_{k}^{\ast}(\tau) \, g_{k}(\tau) = i. 
\label{TWOA3}
\end{equation}
In the discussion of sections \ref{sec3} and \ref{sec4} we are going to emphasize 
the roles of the condition (\ref{TWOA3}) that must always be satisfied throughout all the 
stages of the dynamical evolution. In case the initial vacuum state is annihilated by  $\widehat{a}_{\vec{k},\, \alpha}$
the two-point functions of the field operators and of their time derivatives is:
\begin{eqnarray}
&& \langle \widehat{h}_{i\,j}(\vec{k},\tau) \, \widehat{h}_{m\,n}(\vec{p},\tau) \rangle = \frac{ 2 \pi^2}{k^3} 
\, P_{T}(k,\tau) \,{\mathcal S}_{i\,j\,m\,n}(\hat{k})\, \delta^{(3)}(\vec{k} + \vec{p}), 
\label{TWOA4}\\
&&  \langle \partial_{\tau} \widehat{h}_{i\,j}(\vec{k},\tau) \, \partial_{\tau} \widehat{h}_{m\,n}(\vec{p},\tau) \rangle = \frac{ 2 \pi^2}{k^3} 
\, Q_{T}(k,\tau) \,{\mathcal S}_{i\,j\,m\,n}(\hat{k})\, \delta^{(3)}(\vec{k} + \vec{p}), 
\label{TWOA5}
\end{eqnarray}
where  $P_{T}(k,\tau)$ and $Q_{T}(k,\tau)$ are the two power spectra that depend, respectively, upon 
$F_{k}(\tau)$ and $G_{k}(\tau)$:
\begin{equation}
P_{T}(k,\tau) = \frac{4 \, \ell_{P}^2\, k^3}{\pi^2} \bigl|F_{k}(\tau) \bigr|^2, \qquad 
Q_{T}(k,\tau) = \frac{4 \, \ell_{P}^2\, k^3}{\pi^2} \bigl|G_{k}(\tau) \bigr|^2.
\label{TWOA6}
\end{equation}
While Eq. (\ref{TWOA6}) defines the tensor power spectrum that is often employed in the analysis 
of the conventional inflationary scenarios, the energy and the momentum of the gravitational field
cannot be generally localized and assuming the equivalence principle in its stronger formulation it is conceptually difficult to
propose a unique, local and gauge-invariant definition of the gravitational
energy density \cite{LL,BH,IS} (see also \cite{BG,HL1,HL2}). This problem is particularly acute in the case of the relic gravitons since 
different strategies to assign the energy-momentum pseudo-tensor lead to a variety of problems especially 
for typical wavelengths larger than the Hubble radius \cite{mgen}. Although there exist different and competing 
expressions for the energy-momentum pseudo-tensor of the gravitational field (see e.g. \cite{LL,BH,IS,BG,HL1,HL2}),
 not all of them satisfy the physical requirements that should be reasonably imposed 
 in the case of the relic gravitons \cite{mgen}. It follows however that a consistent 
 expression for the energy-momentum tensor of the gravitational field can be derived 
 from the variation of the second-order action \cite{AC1,AC2} with respect to the background geometry \cite{mgen}
\begin{equation}
T_{\mu}^{(gw)\, \nu} = \frac{1}{4 \,\ell_{P}^2 \, a^2} \biggl[ \partial_{\mu} h_{i\,j} \, \partial^{\nu} h^{i\,j} 
- \frac{1}{2} \,\,\delta_{\mu}^{\nu}\,\, \overline{g}^{\alpha\beta} \,\partial_{\alpha}  h_{i\,j} \partial_{\beta} h^{i\,j} \biggr],
\label{TWOA7}
\end{equation}
as suggested long ago by Ford and Parker \cite{AC1}. From Eq. (\ref{TWOA7}) we can then deduce the energy density of the relic gravitons which is given by:
\begin{equation}
\rho_{gw} = \frac{1}{8 \, \ell_{P}^2 a^2} \biggl[ \partial_{\tau} h_{i\,j} \partial_{\tau} h^{i\,j}  + 
\partial_{k} h_{i\,j} \partial^{k} h^{i\,j} \biggr].
\label{TWOA8}
\end{equation}
If the result of Eq. (\ref{TWOA8}) is now averaged with the help of Eqs. (\ref{TWOA4})--(\ref{TWOA5})
we can get to the explicit expression of the spectral energy density in critical units
\begin{equation}
\Omega_{gw}(k,\tau) = \frac{1}{\rho_{crit}}\frac{d \langle \rho_{gw} \rangle}{d \ln{k}} = \frac{1}{24\, H^2 \,a^2} \biggl[ k^2 P_{T}(k,\tau) + Q_{T}(k,\tau)\biggr],
\label{TWOA9}
\end{equation} 
where $\rho_{crit} = 3\, H^2 \overline{M}_{P}^2 \equiv 3 H^2/\ell_{P}^2$ denotes the critical energy density.
It should be mentioned, at this point, that there are other variables that are customarily employed for the description 
of the random backgrounds of gravitational radiation such as the chirp amplitude $h_{c}(k,\tau)$ or the spectral amplitude 
conventionally denoted by $S_{h}(k,\tau)$. While the chirp amplitude 
can be safely employed in all physical situations, we should stress that 
the diffuse backgrounds of cosmological origin are non stationary \cite{mgstoc}.
The lack of stationarity is ultimately reflected into the spectral energy density and in all the other
observables adopted for the description of a relic signal. A relevant 
consequence of this class of observations is that the spectral amplitude cannot be used for a rigorous
description of the signal. The spectral amplitude (determined, according to the Wiener-Khintchine theorem, by the Fourier transform of the autocorrelation function of the process)
 should be time-independent but, in the case of cosmic gravitons, it is not\footnote{While these issues have been recently discussed at length in \cite{mgstoc}, for the present ends 
 the spectral energy density in critical units and  the chirp amplitude will be the preferred variables;
 we shall therefore avoid using the spectral amplitude.}.

\renewcommand{\theequation}{3.\arabic{equation}}
\setcounter{equation}{0}
\section{Essentials of the WKB evaluation}
\label{sec3}
Even if not strictly necessary, for a consistent WKB solution it is useful to recast Eq. (\ref{TWOA}) (subjected to the condition  (\ref{TWOA3})) in a decoupled form:
\begin{equation}
f_{k}^{\prime\prime} + [ k^2 - a^{\prime\prime}/a] f_{k} =0, \qquad 
g_{k}= f_{k}^{\prime} - {\mathcal H} f_{k}.
\label{WKB1}
\end{equation}
For $k^2 \gg |a^{\prime\prime}/a|$ the solutions of Eq. (\ref{WKB1}) are plane waves and, depending on the shape of $|a^{\prime\prime}/a|$, this regime may also correspond to the limits $\tau \to \pm \infty$ but this correspondence is not strictly necessary, at least in the first approximation. Conversely, 
for $k^2 \ll |a^{\prime\prime}/a|$ the solution of Eq. (\ref{WKB1}) 
can be obtained by first transforming Eq. (\ref{WKB1}) into an integral 
equation 
\begin{equation}
f_{k}(\tau) = f_{k}(\tau_{ex}) \frac{a(\tau)}{a_{ex}} + g_{k}(\tau_{ex}) \frac{a(\tau)}{a_{ex}}\, \int_{\tau_{ex}(k)}^{\tau} \frac{a_{ex}^2}{a^2(\tau^{\prime})} \, d\tau^{\prime}
- k^2 \, a(\tau) \, \int_{\tau_{ex}(k)}^{\tau} \frac{d\,\tau^{\prime}}{a^2(\tau^{\prime})} \int_{\tau_{ex}(k)}^{\tau^{\prime}} \, a(\tau^{\prime\prime}) \, f_{k}(\tau^{\prime\prime})\, d\, \tau^{\prime\prime},
\label{WKB2}
\end{equation}
where $\tau_{ex}(k)$ indicates the time when the wavelength $2\pi/k$ exits the Hubble radius (during the inflationary stage); in practice $\tau_{ex}(k)$ is determined from the (approximate) equality $k^2 \simeq  |a^{\prime\prime}/a|_{\tau_{ex}}$. Equation (\ref{WKB2}) can be iteratively solved and, for instance, the first 
iteration consists in neglecting the term that contains $k^2$ that are 
effectively subleading when the relevant wavelengths are larger than the Hubble radius. The 
value of $g_{k}(\tau)$ can be either computed from $g_{k}= (f_{k}^{\prime} - {\mathcal H} \, f_{k})$ through Eq. (\ref{WKB2}) or 
it can be directly obtained from the decoupled evolution of $g_{k}(\tau)$, i.e. 
$g_{k}^{\prime\prime} + [ k^2 - a (1/a)^{\prime\prime}] g_{k} =0$. In both approaches the values of 
$f_{k}(\tau_{ex})$ and $g_{k}(\tau_{ex})$ must comply with the Wronskian normalization condition (\ref{TWOA3}) applied at 
$\tau_{ex}(k)$, i.e. 
\begin{equation}
f_{k}(\tau_{ex}) \,g^{\ast}_{k}(\tau_{ex}) - f_{k}^{\ast}(\tau_{ex}) \,g_{k}(\tau_{ex}) = i. 
\label{WKB3}
\end{equation}
Finally, the solutions  in the regime $k^2 \gg  |a^{\prime\prime}/a|$ and $k^2 \ll  |a^{\prime\prime}/a|$
can be matched at $\tau_{ex}(k)$ and this procedure leads to a sound approximation 
for typical wavelengths larger than the Hubble radius. 
 
\subsection{The mode functions in the WKB approach}
We are ultimately interested 
in the solution valid for wavelengths that have already reentered the Hubble radius at the present time. 
Indeed the validity of the approximation scheme leading to Eq. (\ref{WKB2}) stops 
at $\tau_{re}(k)$, i.e. at the approximate time scale where a given wavelength reenters 
the Hubble radius; the solutions for the mode functions in the 
range $\tau> \tau_{re}(k)$ are again standing waves and can be written in the form: 
\begin{eqnarray}
f_{k}(\tau) &=& \frac{e^{- i k \tau_{ex}}}{\sqrt{ 2 k}} \biggl[ {\mathcal M}_{k}(\tau_{ex},\tau_{re}) \, \cos{k \,\Delta\tau} + {\mathcal N}_{k}(\tau_{ex},\tau_{re}) \, \sin{k \,\Delta\tau} \biggr],
\label{WKB4}\\
g_{k}(\tau) &=& e^{- i k \tau_{ex}}\, \sqrt{\frac{k}{2}}\, \biggl[ - {\mathcal M}_{k}(\tau_{ex},\tau_{re}) \, \sin{k \,\Delta\tau} + \mathcal{N}_{k}(\tau_{ex},\tau_{re}) \, \cos{k \,\Delta\tau} \biggr],
\label{WKB5}
\end{eqnarray}
where $\Delta \tau = [\tau - \tau_{re}(k)]$. The explicit form of the functions ${\mathcal M}_{k}(\tau_{ex},\tau_{re})$ and ${\mathcal N}_{k}(\tau_{ex},\tau_{re})$ follows from the approximate solution (\ref{WKB2}) together with the requirement that the initial conditions in the limit $\tau \to - \infty$ are represented by travelling waves that obey the Wronskian normalization condition. The approximate solutions must then be continuously matched in $\tau_{ex}(k)$ and $\tau_{re}(k)$ by always bearing in mind 
the condition (\ref{WKB3}) which also constrain the explicit form of ${\mathcal M}_{k}(\tau_{ex},\tau_{re}) $ and 
${\mathcal N}_{k}(\tau_{ex},\tau_{re})$.  The result of this lengthy but straightforward procedure is:
\begin{eqnarray}
{\mathcal M}_{k}(\tau_{ex},\tau_{re}) &=& \biggl(\frac{a_{re}}{a_{ex}}\biggr) \, Q_{k}(\tau_{ex}, \tau_{re}),
\label{WKB6}\\
{\mathcal N}_{k}(\tau_{ex}, \tau_{re}) &=& \biggl(\frac{{\mathcal H}_{re}}{k} \biggr) \,  \biggl(\frac{a_{re}}{a_{ex}}\biggr) \, Q_{k}(\tau_{ex}, \tau_{re})
- \biggl(\frac{a_{ex}}{a_{re}}\biggr) \biggl(\frac{{\mathcal H}_{ex} + i k}{k}\biggr),
\label{WKB7}\\
Q_{k}(\tau_{ex}, \tau_{re}) &=& 1 - ({\mathcal H}_{ex} + i k) \int_{\tau_{ex}}^{\tau_{re}} \frac{a_{ex}^2}{a^2(\tau^{\prime})} \,\, d \tau^{\prime},
\label{WKB8}
\end{eqnarray}
where $a_{re} = a(\tau_{re})$, $a_{ex} = a(\tau_{ex})$ while the other quantities, with obvious notations, are:
\begin{equation}
{\mathcal H}_{re} = {\mathcal H}(\tau_{re})=  a_{re} H_{re}, \qquad {\mathcal H}_{ex} = {\mathcal H}(\tau_{ex})= a_{ex} H_{ex},\qquad
{\mathcal I}(\tau_{ex},\tau_{re}) = \int_{\tau_{ex}(k)}^{\tau_{re}(k)} \frac{a_{ex}^2}{a^2(\tau)} \, d\tau.
\end{equation}
It can be directly checked that the Wronskian 
normalization condition (\ref{TWOA3}) is always satisfied by Eqs. (\ref{WKB4})--(\ref{WKB5}) 
provided ${\mathcal M}_{k}(\tau_{ex},\tau_{re})$ and ${\mathcal N}_{k}(\tau_{ex},\tau_{re})$ have exactly the form given in Eqs. (\ref{WKB6})--(\ref{WKB7}) and (\ref{WKB8}).  If Eqs. (\ref{WKB4})-(\ref{WKB5}) are inserted into Eqs. (\ref{TWOA5})--(\ref{TWOA6}) and (\ref{TWOA9}) it turns out that the spectral energy density in critical units depends on $|{\mathcal M}_{k}(\tau_{ex},\tau_{re})|^2$ and $|{\mathcal N}_{k}(\tau_{ex},\tau_{re})|^2$:
\begin{equation}
\Omega_{gw}(k,\tau) = \frac{2 \, k^4}{3 \, \pi \, H^2\, a^4\, M_{P}^2} \biggl[ 
\biggl| {\mathcal M}_{k}(\tau_{ex},\tau_{re}) \biggr|^2 + \biggl| {\mathcal N}_{k}(\tau_{ex},\tau_{re}) \biggr|^2\biggr].
\label{WKB9}
\end{equation}
If we now insert Eqs. (\ref{WKB6})--(\ref{WKB7}) into Eq. (\ref{WKB9}) the explicit 
form of $\Omega_{gw}(k,\tau)$ becomes:
\begin{eqnarray}
\Omega_{gw}(k,\tau) &=& \frac{2 \, k^4}{3 \, \pi \, H^2\, a^4\, M_{P}^2}\biggl\{ \biggl(\frac{a_{re}}{a_{ex}}\biggr)^2 \biggl( 1 + \frac{{\mathcal H}_{re}^2}{k^2}\biggr) \biggl[ 1 + ({\mathcal H}_{ex}^2 +k^2) {\mathcal I}^2(\tau_{ex},\tau_{re}) - 2 {\mathcal H}_{ex} {\mathcal I}(\tau_{ex},\tau_{re})\biggr] 
\nonumber\\
&-&  2 \frac{{\mathcal H}_{re}}{k}\biggl[ \frac{{\mathcal H}_{ex}}{k} - \frac{{\mathcal H}_{ex}^2 + k^2 }{k}\, {\mathcal I}(\tau_{ex},\tau_{re})\biggr]+\biggl(\frac{a_{ex}}{a_{re}}\biggr)^2 \biggl[ 1 + \frac{{\mathcal H}_{ex}^2}{k^2}\biggr]\biggr\}.
\label{WKB10}
\end{eqnarray}
In connection with Eq. (\ref{WKB10}) there are two important observations.
The first one is that  ${\mathcal H}_{ex} \simeq k$ at least as long as the exit  of the given wavelength occurs during an inflationary stage of expansion (see however some exceptions described at the end of this subsection). The second observation is that there is a hierarchy between the first and the last 
terms of Eq. (\ref{WKB10}) implying that the first term always dominates against the last one: if 
the scale factor expands $a_{re} \gg a_{ex}$ so that the term containing $(a_{re}/a_{ex})^2$ 
is always negligible for all practical purposes. Thanks to these two remarks Eq. (\ref{WKB10}) 
can be rewritten as:
\begin{eqnarray}
\Omega_{gw}(k,\tau) &=& \frac{2 \, k^4}{3 \, \pi \, H^2\, a^4\, M_{P}^2}\biggl\{ \biggl(\frac{a_{re}}{a_{ex}}\biggr)^2 \biggl( 1 + \frac{{\mathcal H}_{re}^2}{k^2}\biggr) \biggl[ 1 + 2 \,k^2\, {\mathcal I}^2(\tau_{ex},\tau_{re}) 
\nonumber\\
&-& 2 \, k\,{\mathcal I}(\tau_{ex},\tau_{re})\biggr] - 2 \frac{{\mathcal H}_{re}}{k}\biggl[ 1 - 2 \, k\, {\mathcal I}(\tau_{ex},\tau_{re})\biggr]  \biggr\}.
\label{WKB11}
\end{eqnarray}
The values of $\tau_{ex}(k)$ and $\tau_{re}(k)$ define the change of the solution and are therefore the two turning points of the WKB approximation. From Eq. (\ref{WKB1}) the turning points approximately 
obey $k^2 \simeq |a^{\prime\prime}/a|$. The latter condition can also be written 
as $k^2 \simeq a^2 \, H^2 [ 2 - \epsilon]$ where $\epsilon = - \dot{H}/H^2$ and 
since during inflation $\epsilon \ll 1$ the turning point is always regular and, as 
anticipated, $k\simeq a_{ek} H_{ex} = {\mathcal H}_{ex}$.  It can however happen that the  reentry occurs either during the radiation stage or close to it: in this case  $ \epsilon= {\mathcal O}(2)$ and this implies that 
$ k\ll a_{re}\, H_{re} = {\mathcal H}_{re}$. Thus when the reentry  takes place during a radiation stage the terms ${\mathcal H}_{re}/k \gg 1$ dominate in Eq. (\ref{WKB11}). 

\subsection{Integrals for wavelengths larger than the Hubble radius}
Within the WKB approach the high frequency normalization does not only depend upon the frequency slope but also 
on the evaluation of certain integrals whose numerical relevance may range between 
few percents and one order of magnitude.  Although these contributions could simply be disregarded, for the sake of accuracy (and for a 
fair comparison with the results of section \ref{sec5}) we are now going to evaluate the integrals appearing in Eq. (\ref{WKB11}) by bearing in mind that they all involve a physical regime where the corresponding wavelengths are larger than the Hubble radius. Since for $\tau \to - \infty$ we assume an inflationary stage of expansion the (comoving) Hubble rate can be written as:
\begin{equation}
H\, a = - \frac{1}{(1 - \epsilon) \, \tau}, \qquad \epsilon = - \dot{H}/H^2\ll 1, \qquad \tau \leq - \tau_{1}.
\label{TWOB}
\end{equation}
We then posit that for $\tau \geq - \tau_{1}$ there is a generic stage of decelerated expansion (i.e. $\delta > 0$)
\begin{equation}
H\, a = \frac{\delta}{ \tau + \tau_{1} ( 1 + q) }, \qquad q = q(\epsilon, \delta)= (1 - \epsilon)\delta, \qquad \tau \geq - \tau_{1}.
\label{TWOC}
\end{equation}
If $\delta \to 1$ the background is dominated by radiation; conversely when $\delta \neq 1$) the expansion rate is, respectively, either 
slower (i.e. $\delta < 1$) or faster (i.e. $\delta > 1$) than radiation. Because of the presence of $q(\epsilon, \delta)$ in Eqs. (\ref{TWOB})--(\ref{TWOC})  
the scale factor and the expansion rate are explicitly continuous across the transition for any value of $\delta$ and $\epsilon$. The integral $k\,{\mathcal I}(\tau_{ex}, \tau_{re})$ can then be separated into the two 
contributions preceding and following $- \tau_{1}$, namely:
\begin{equation}
k\,{\mathcal I}(\tau_{ex}, \tau_{re}) = k \, \int_{-1/k}^{-\tau_{1}} \frac{a_{ex}^2}{a^2(\tau)} d\tau + 
k  \int_{-\tau_{1}}^{1/k} \frac{a_{ex}^2}{a^2(\tau)} d\tau.
\label{INT1}
\end{equation}
Since the exit always occurs during the inflationary stage we can estimate $a_{ex}$ as $a_{ex}^2 = (- \tau_{ex}/\tau_{1})^{-2/(1 - \epsilon)} \simeq |k \tau_{1}|^{2/(1- \epsilon)}$; therefore we have from Eq. (\ref{INT1}) that 
\begin{equation}
k\,{\mathcal I}(\tau_{ex}, \tau_{re}) = \frac{1}{p(\epsilon)} \biggl[ 1 - x_{1}^{p(\epsilon)}\biggr] + q^{2 \delta} x_{1}^{2 \delta + 2/(1 - \epsilon)} \int_{- x_{1}}^{1} 
\frac{d x}{[ x + (q + 1) x_{1}]^{2 \delta}},
\label{INT2}
\end{equation}
where the auxiliary quantity $p(\epsilon) =(3 - \epsilon)/(1 - \epsilon)$ has been introduced.
Provided $\delta \neq 1/2$ we have that the whole integral appearing in Eq. (\ref{INT2}) becomes
\begin{eqnarray}
k\,{\mathcal I}(\tau_{ex}, \tau_{re}) &=&\frac{1}{p(\epsilon)} \biggl[ 1 - x_{1}^{p(\epsilon)}\biggr] + \frac{q^{2 \delta} x_{1}^{2 \delta + 2/(1 - \epsilon)}}{1 - 2 \delta} \biggl\{ \biggl[ 1 + \biggl(q +1\biggr)x_{1}\biggr]^{1 - 2 \delta} - \biggl(q \,x_{1}\biggr)^{1 - 2\delta}\biggr\} 
\nonumber\\
&=& \frac{1}{p(\epsilon)}  + {\mathcal O}\biggl(x_{1}^{3 + 2 \epsilon}\biggr) + {\mathcal O}\biggl(x_{1}^{2 \delta + 2 + 2 \epsilon}\biggr).
\label{INT3}
\end{eqnarray}
In Eq. (\ref{INT3}) all the corrections are subleading as long as $x_{1} <1$; this happens for all the wavenumbers smaller than the maximal one and for all the frequencies smaller than the maximal frequency. In the case $ \delta \to 1/2$ the evaluation is similar but the final result contains logarithmic corrections that can be estimated as follows:
\begin{equation}
k\,{\mathcal I}(\tau_{ex}, \tau_{re}) =  \frac{1}{p(\epsilon)}- q \, x_{1}^{p(\epsilon)}\, \ln{(q\, x_{1})}.
\label{INT4}
\end{equation}
We can now go back to Eq. (\ref{WKB11}) and note that the first term is always 
dominant in the case of an expanding background since it is proportional to $(a_{re}/a_{ex})\gg 1$.
This means that Eq. (\ref{WKB11}) can also be written as:
 \begin{equation}
\Omega_{gw}(k,\tau) = \frac{2 \, k^4}{3 \, \pi \, H^2\, a^4\, M_{P}^2}\biggl(\frac{a_{re}}{a_{ex}}\biggr)^2 \, \biggl( 1 + \frac{{\mathcal H}_{re}^2}{k^2}\biggr) \biggl[ 1 + 2 \,k^2\, {\mathcal I}^2(\tau_{ex},\tau_{re}) - 2 \, k\,{\mathcal I}(\tau_{ex},\tau_{re})\biggr].
\label{WKB11a}
\end{equation}

\subsection{The spectral energy density for different post-inflationary histories}
The explicit expression of the spectral energy density will now be examined for different 
post-inflationary histories within the approximation schemes studied in this section.
We first consider the situation where the given wavelength reenters the Hubble radius 
far from a radiation-dominated stage of expansion (i.e. $\epsilon_{re} \neq 2$); we furthermore 
require that $\delta \neq 1/2$ to avoid logarithmic corrections. From Eq. (\ref{WKB11a}) we have 
\begin{equation}
\Omega_{gw}(k,\tau) = \frac{2 \, k^4 c_{1}(\epsilon)}{3 \, \pi \, H^2\, a^4\, M_{P}^2}\biggl(\frac{a_{re}}{a_{ex}}\biggr)^2,
\label{WKB11b}
\end{equation}
where the auxiliary constant $c_{1}(\epsilon)$ has been introduced and it is defined as:
\begin{equation}
 c_{1}(\epsilon) =2 \biggl[ 1 + \frac{2}{p^2(\epsilon)} - \frac{2}{p(\epsilon)}\biggr] =2 \frac{[4 + (1 - \epsilon)^2]}{(3 - \epsilon)^2}. 
 \label{WKB11bb}
\end{equation}
As anticipated this factor comes from the integrals evaluated above but its actual numerical weight 
is ${\mathcal O}(1)$: in the limit $\epsilon \to 0$ we have that $c_{1} \to 10/3$.
We must then estimate explicitly $(a_{re}/a_{ex})^2$ in terms of Eqs. (\ref{TWOB})--(\ref{TWOC}) and the result is:
\begin{equation}
\frac{a_{re}^2}{a_{ex}^2} = \frac{[(\tau_{re}/\tau_{1} +1)/q +1]^{2 \delta}}{(\tau_{ex}/\tau_{1})^{- 2/(1 - \epsilon)}} = q^{-2 \delta} \, \biggl(\frac{k}{a_{1} \, H_{1}}\biggr)^{- 2\delta - 2/(1 - \epsilon)}.
\label{WKB11c}
\end{equation}
For a concrete estimate of the spectral energy density of Eq. (\ref{WKB11b}) we consider that the $\delta$-phase of Eq. (\ref{TWOC}) extends between $H_{1}$ down to a curvature scale $H_{r}$ marking the onset of the radiation-dominated stage. When $H < H_{r}$ the standard timeline of the concordance paradigm suggests that
$ 3 H_{eq}^2 \overline{M}_{P}^2 = 2 \rho_{M0}(a_{0}/a_{eq})^3$ at the time of matter-radiation 
equality so that\footnote{We recall that $\Omega_{R\,0}$ has been already introduced in Eq. 
(\ref{FF2}) while $\Omega_{M0}$ is the critical fraction of matter in the concordance scenario} $(a_{0}/a_{eq}) = \Omega_{M\,0}/\Omega_{R\,0}$ where; with the same logic we can also
compute  $(a_{0} \, H_{0})/(a_{eq}\, H_{eq}) = (2 \, \Omega_{R\, 0})^{-1/4} \sqrt{H_{0}/H_{eq}}$.
Since between $H_{r}$ and $H_{eq}$  the evolution is dominated by radiation we also have that
\begin{equation}
\biggl(\frac{a_{eq}^4 \, H_{eq}^2}{a_{r}^4 H_{r}^2}\biggr) = \frac{a_{eq}^4 \, T_{eq}^4 \, g_{\rho,\,eq}}{a_{r}^4 \, T_{r}^4 g_{\rho,\, r}} = \biggl(\frac{g_{\rho,\,eq}}{g_{\rho,\,r}}\biggr)  \biggl(\frac{g_{s,\,r}}{g_{s,\,eq}}\biggr)^{4/3}.
\label{GG2}
\end{equation} 
In Eq. (\ref{GG2}) $g_{\rho}$ and $g_{s}$ denote, respectively, the number of effective relativistic degrees of freedom appearing in the energy and entropy densities of the plasma. This simplified estimate follows by assuming local thermal 
equilibrium at $H_{r}$; thus if the entropy density is conserved $g_{s,\,r} \, a_{r}^3 \, T_{r}^3 = g_{s,\,eq} \, a_{eq}^3 \, T_{eq}^3$ and this observation determines the redshift between the two epochs. In the standard situation where $g_{s,\, r}= g_{\rho,\, r} = 106.75$ and $g_{s,\, eq}= g_{\rho,\, eq} = 3.94$ so that difference due to $g_{s}$ and $g_{\rho}$ in the final results is actually negligible for the present purposes\footnote{The difference due to $g_{s}$ and $g_{\rho}$ in the final results is actually involves a factor $1.3$ (instead of $1$) at the level of Eq. (\ref{GG2}).}. Thanks to Eqs. (\ref{WKB11b})--(\ref{WKB11c}) and (\ref{GG2}) we can therefore obtain the wanted estimate of $\Omega_{gw}(k,\tau)$ in the WKB approximation
\begin{equation}
\Omega_{gw}(k,\tau) = \frac{4 \,\Omega_{R\,0} \, c_{1}(\epsilon)}{3 \pi \, q^{2 \delta}} \,\biggl(\frac{H_{1}}{M_{P}}\biggr)^2 \, \biggl(\frac{H_{r}}{H_{1}}\biggr)^{\alpha(\delta)} \biggl(\frac{g_{\rho,\,r}}{g_{\rho,\,eq}}\biggr)  \biggl(\frac{g_{s,\,eq}}{g_{s\,r}}\biggr)^{4/3} \biggl(\frac{k}{a_{1}\, H_{1}}\biggr)^{n_{T}(\delta,r_{T})}.
\label{OMWK1}
\end{equation}
In Eq. (\ref{OMWK1}) the spectral index $n_{T}(\delta, r_{T})$ controls the slope of the spectral energy density while 
$\alpha(\delta)$ affects the overall amplitude of $\Omega_{gw}(k,\tau)$; 
 the spectral index $n_{T}(\delta, r_{T})$ and $\alpha(\delta)$ are given, respectively, by 
\begin{equation}
n_{T}(\delta,r_{T})= 4 - \frac{32}{16 - r_{T}} - 2 \delta, \qquad\qquad \alpha(\delta) = \frac{2 (\delta -1)}{\delta + 1}.
\label{SPIN1}
\end{equation}
The result of Eq. (\ref{OMWK1}) can be made even more explicit 
by recalling that $(H_{1}/M_{P})= \sqrt{ \pi \, {\mathcal A}_{{\mathcal R}} \, r_{T}}/4$:
\begin{equation}
\Omega_{gw}(k,\tau) = \frac{r_{T}\, {\mathcal A}_{{\mathcal R}}\,  \Omega_{R\,0} \, c_{1}(\epsilon)}{12 \, q^{2 \delta}} \, \biggl(\frac{H_{r}}{H_{1}}\biggr)^{\alpha(\delta)} \biggl(\frac{g_{\rho,\,r}}{g_{\rho,\,eq}}\biggr)  \biggl(\frac{g_{s,\,eq}}{g_{s\,r}}\biggr)^{4/3} \biggl(\frac{k}{a_{1}\, H_{1}}\biggr)^{n_{T}(\delta,r_{T})},
\label{OMWK2}
\end{equation}
where ${\mathcal A}_{{\mathcal R}}= {\mathcal O}(10^{-9})$ denotes the amplitude of the scalar (i.e. curvature) inhomogeneities at the pivot scale $k_{p}$ \cite{SW1,TT1,TT2,TT3}. The results of Eqs. (\ref{OMWK1}) and (\ref{OMWK2}) have been deduced in the case $\delta \neq 1/2$ but when $\delta \to 1/2$ the leading term is formally the same with the difference that $c_{1}(\epsilon, k)$ 
is now scale dependent:
\begin{equation}
\Omega_{gw}(k,\tau) = \frac{r_{T}\, {\mathcal A}_{{\mathcal R}}\,  \Omega_{R\,0} \, c_{1}(\epsilon,\,k)}{12 \, q^{2 \delta}} \, \biggl(\frac{H_{r}}{H_{1}}\biggr)^{\alpha(\delta)} \biggl(\frac{g_{\rho,\,r}}{g_{\rho,\,eq}}\biggr)  \biggl(\frac{g_{s,\,eq}}{g_{s\,r}}\biggr)^{4/3} \biggl(\frac{k}{a_{1}\, H_{1}}\biggr)^{n_{T}(1/2,r_{T})}, 
\label{OMWK2a}
\end{equation}
Two further auxiliary constants have been introduced in Eq. (\ref{OMWK2a}) and are defined as defined as:
\begin{eqnarray}
c_{1}(\epsilon, k) &=&  2 \frac{[4 + (1 - \epsilon)^2]}{(3 - \epsilon)^2}\biggl[ 1 + c_{2}(\epsilon) \, q\, \biggl(\frac{k}{a_{1}\, H_{1}}\biggr)^{p(\epsilon)} \ln{\biggl(\frac{k}{a_{1}\, H_{1}}\biggr)}\biggr], 
\nonumber\\
c_{2}(\epsilon) &=& \frac{2 ( 3 + 2\epsilon - \epsilon^2)}{5 - 2 \epsilon + \epsilon^2}.
\label{OMWK2aa}
\end{eqnarray}
To complete the discussion we can finally deduce the spectral energy density 
when the reentry takes place in the radiation stage while the exit 
occurs, as usual, during an inflationary phase. In this situation we have 
that the conditions $\epsilon_{re} = {\mathcal O}(2)$ and $\epsilon_{ex} = \epsilon \ll 1$
imply that the exit is a regular turning point (i.e. $k \simeq a_{ex} \, H_{ex}$)
 while at reentry we would have instead  $ k^2 = a_{re}^2 \, H_{re}^2 [ 2 - \epsilon_{re}]$, i.e. $k \ll a_{re} H_{re}$. As already anticipated, the term 
${\mathcal H}_{re}^2/k^2 \gg 1$ and from Eq. (\ref{WKB11b}) we obtain: 
\begin{equation}
\Omega_{gw}(k,\tau) = \frac{2 \, k^2 a_{re}^2 \, H_{re}^2\, c_{3}(\epsilon)}{3 \, \pi \, H^2\, a^4\, M_{P}^2}\biggl(\frac{a_{re}}{a_{ex}}\biggr)^2,   
\label{OMWK2b}
\end{equation}
where $c_{3}(\epsilon) = c_{2}(\epsilon)/2$ and in the limit $\epsilon\to 0$ $c_{2} \to 6/5$. The dominant contribution written in Eq. (\ref{OMWK2b}) now follows from ${\mathcal H}_{re}^2/k^2 \gg 1$; for a more direct evaluation  $\Omega_{gw}(k,\tau)$ can be written as:
\begin{equation}
\Omega_{gw}(k,\tau) = \frac{2 \,  c_{3}(\epsilon)}{3 \, \pi} \biggl(\frac{k^2}{a_{ex}^2 \, H_{ex}^2}\biggr) \biggl(\frac{H_{ex}}{M_{P}}\biggr)^2 \frac{a_{re}^4 \, H_{re}^2}{a^4 \, H^2},
\label{OMWK2c}
\end{equation}
where the right-hand side has been multiplied and divided by $H_{ex}^2$.
Since $k \simeq a_{ex} \, H_{ex}$,  Eq. (\ref{OMWK2c})  further simplifies: the second term is of order $1$ while $H_{ex}^2 = H_{1}^2 |k/(a_{1}\, H_{1})|^{\overline{n}_{T}}$ where, in this case, we denote $\overline{n}_{T} = - 2\epsilon/(1- \epsilon) \simeq - 2 \epsilon$. Since the consistency relations have been enforced throughout the discussion $\overline{n}_{T} = - 2 \epsilon \simeq - r_{T}/8$. We then have, from Eq. (\ref{OMWK2c}), that the estimate of spectral energy density is consistent with the limit 
of Eq. (\ref{OMWK2a}) for $\delta \to 1$
\begin{equation}
\Omega_{gw}(k,\tau) =
 \frac{r_{T}\, {\mathcal A}_{{\mathcal R}}\,  \Omega_{R\,0} \, c_{3}(\epsilon)}{12} \, \biggl(\frac{g_{\rho,\,r}}{g_{\rho,\,eq}}\biggr)  \biggl(\frac{g_{s,\,eq}}{g_{s\,r}}\biggr)^{4/3} \biggl(\frac{k}{a_{1}\, H_{1}}\biggr)^{\overline{n}_{T}},\qquad \overline{n}_{T} \simeq - r_{T}/8,
\label{OMWK2d}
\end{equation}
both at the level of the amplitude and of the spectral index. Indeed from Eq. (\ref{SPIN1}) we have that 
\begin{equation}
\lim_{\delta\to 1,\,\,r_{T} \ll 1} n_{T}(\delta, r_{T}) = \lim_{r_{T} \ll 1} \biggl(2 - \frac{32}{16 - r_{T}}\biggr) \to  - r_{T}/8,
\end{equation}
and it coincides with $\overline{n}_{T}$ appearing in Eq. (\ref{OMWK2d}).  In this analysis a single post-inflationary stage between $H_{1}$ and $H_{r}$ has been assumed but the obtained results can be easily generalized to the situation where a single phase is replaced by a series of different epochs\footnote{
The spectral energy density of Eq. (\ref{OMWK1}) will then have more than one high frequency branch with different spectral indices characterizing each spectral domain. These possible extensions 
are not examined here but our results equally apply to a multi-stage post-inflationary evolution previously analyzed in Refs. \cite{MAX1,MAX2}}.

\subsection{The maximal frequency}
While deriving the explicit form of the spectral energy density we assumed throughout that $a_{0}=1$; this condition implies that at the present time comoving and physical wavenumbers coincide. Bearing in mind this point we can express more 
explicitly $k/(a_{1} \, H_{1})$:
\begin{equation}
\frac{k}{a_{1} H_{1}} = \frac{k}{a_{0} \, H_{0}} \biggl(\frac{a_{0} \, H_{0}}{a_{eq} \, H_{eq}}\biggr) \biggl(\frac{a_{eq} \, H_{eq}}{a_{r} \, H_{r}}\biggr) \biggl(\frac{a_{r} \, H_{r}}{a_{1} \, H_{1}}\biggr).
\label{kx1}
\end{equation}
By now recalling that $k = 2 \pi \nu$ we can therefore obtain the spectral energy density 
at the present time as a function of the comoving frequencies $\nu$ and $\nu_{max}$ since $x_{1} = (\nu/\nu_{max})$. For instance, from Eq. (\ref{OMWK2}) we have 
\begin{equation}
\Omega_{gw}(\nu,\tau_{0}) = \frac{r_{T}\, {\mathcal A}_{{\mathcal R}}\,  \Omega_{R\,0} \, c_{1}(\epsilon)}{12 \, q^{2 \delta}} \, \biggl(\frac{H_{r}}{H_{1}}\biggr)^{\alpha(\delta)} \biggl(\frac{g_{\rho,\,r}}{g_{\rho,\,eq}}\biggr)  \biggl(\frac{g_{s,\,eq}}{g_{s\,r}}\biggr)^{4/3} \biggl(\frac{\nu}{\nu_{max}}\biggr)^{n_{T}(\delta,r_{T})},
\label{OMWK2c1}
\end{equation}
where the explicit value of the maximal frequency $\nu_{max}$ is now given by:
\begin{equation}
\nu_{max} = \frac{\sqrt{H_{0} \, M_{P}}}{2 \pi} \, (2 \, \Omega_{R\,0})^{1/4} \sqrt{\frac{H_{1}}{M_{P}}}
\biggl(\frac{g_{\rho,\, r}}{g_{\rho,\,eq}}\biggr)^{1/4} \biggl(\frac{g_{s,\, eq}}{g_{s,r}}\biggr)^{1/3} 
\biggl(\frac{H_{r}}{H_{1}}\biggr)^{\alpha(\delta)/4}.
\label{FREQ2}
\end{equation}
The result of Eq. (\ref{FREQ2}) can also be be expressed, with shorthand notation, as 
$\nu_{max} = (H_{r}/H_{1})^{\alpha(\delta)/4}\, \overline{\nu}_{max}$ where 
$\overline{\nu}_{max}$ corresponds to the maximal frequency when the post-inflationary evolution 
is dominated by radiation from $H_{1}$ down to $H_{eq}$. 
In this limit the explicit value of $\overline{\nu}_{max}$ falls in the range of ${\mathcal O}(300)\, \mathrm{MHz}$ and, more precisely, we have 
\begin{equation}
\overline{\nu}_{max} = 271.9 \, \biggl(\frac{h_{0}^2 \, \Omega_{R\, 0}}{4.15\times 10^{-5}}\biggr) 
\biggl(\frac{r_{T}}{0.06}\biggr)^{1/4} \, \biggl(\frac{{\mathcal A}_{{\mathcal R}}}{2.41\times 10^{-9}}\biggr)^{1/4} \, \biggl(\frac{g_{\rho,\, r}}{g_{\rho,\,eq}}\biggr)^{1/4} \biggl(\frac{g_{s,\, eq}}{g_{s,r}}\biggr)^{1/3} \, \mathrm{MHz}.
\label{FREQ3}
\end{equation}
When $g_{s,\, r}= g_{\rho,\, r} = 106.75$ and $g_{s,\, eq}= g_{\rho,\, eq} = 3.94$ the evolution of the relativistic species 
suppresses $\overline{\nu}_{max}$ by a factor ${\mathcal O})(0.75)$; this means that $\overline{\nu}_{max}$ moves from $271.9\, \mathrm{MHz}$ down to $203.9$ MHz. Both in Eqs. (\ref{OMWK2d}) and (\ref{FREQ2}) we simply considered a single post-inflationary stage preceding the radiation-dominated evolution.
The timeline of the expansion rate can be however more complicated 
and there may be different post-inflationary stages for curvature scales larger than $H_{r}$.  In all these situations the bunch of frequencies $\nu = {\mathcal O}(\nu_{max})$ will always correspond to the wavelengths that left the horizon at the end of inflation and reentered immediately after. Depending on the timeline of the post-inflationary evolution there will however be other typical frequencies between below $\nu_{max}$ \cite{MGrev}.  

\renewcommand{\theequation}{4.\arabic{equation}}
\setcounter{equation}{0}
\section{The transition matrix and the exact evaluation}
\label{sec4}
If both $a(\tau)$ and ${\mathcal H}(\tau) = a\, H$ are continuous 
throughout all the stages of the dynamical evolution, at late times the 
mode functions and the spectral energy density can be computed from the elements of a transition matrix that relates the inflationary power spectra to their decelerated counterpart.
From Eqs. (\ref{TWOA}) and (\ref{TWOB}) it follows that during the inflationary 
stage of expansion (see e.g. Eq. (\ref{TWOC})) the 
explicit expressions of $f_{k}(\tau)$ and $g_{k}(\tau)$ are given by\footnote{As usual $H_{\mu}^{(1)}(z)$ denotes the Hankel function of first kind with index $\mu$ and argument $z$; we also recall that, by definition, $H_{\mu}^{(2)}(z)= H_{\mu}^{(1)\ast}(z)$.} :
\begin{eqnarray}
f_{k}(\tau) &=& \frac{N_{\mu}}{\sqrt{2 k}}\, \sqrt{ - k \tau} \, H_{\mu}^{(1)}(- k \tau), \qquad 
\mu = \frac{3 - \epsilon}{2 ( 1 - \epsilon)},
\label{TWOD}\\
g_{k}(\tau) &=& - N_{\mu} \sqrt{\frac{k}{2}} \, \sqrt{- k\,\tau}\, H_{\mu-1}^{(1)}(- k \tau),
\label{TWOE}
\end{eqnarray}
where $N_{\mu} = \sqrt{\pi/2}\, e^{ i \pi ( 2 \mu+1)/4}$ while, as before, $\epsilon= - \dot{H}/H^2$; note that Eqs. (\ref{TWOD})--(\ref{TWOE}) are valid for $\tau \leq - \tau_{1}$.
Owing to the continuity of the background fields it is always possible to relate the mode functions for $ \tau \geq - \tau_{1}$ to the ones defined during inflation and the general form of this relation depends on four coefficients that define the entries of the transition matrix:
\begin{eqnarray}
f_{k}(\tau) &=& {\mathcal C}_{f\,f}(k,\tau,\tau_{1})\, \overline{f}_{k} + {\mathcal C}_{f\,g}(k,\tau,\tau_{1})\,\overline{g}_{k}/k,
\label{THREEA}\\
g_{k}(\tau) &=& {\mathcal C}_{g\,f}(k,\tau,\tau_{1}) k\,\overline{f}_{k} + {\mathcal C}_{g\,g}(k,\tau,\tau_{1}) \, \overline{g}_{k}.
\label{THREEB}
\end{eqnarray}
From Eqs. (\ref{TWOD})--(\ref{TWOE}) we have that $\overline{f}_{k} = f_{k}(-\tau_{1})$ and  $\overline{g}_{k} = g_{k}(-\tau_{1})$. The four functions ${\mathcal C}_{f\,f}$, ${\mathcal C}_{f\,g}$, ${\mathcal C}_{g\,f}$ and ${\mathcal C}_{g\,g}$ appearing in Eqs. (\ref{THREEA})--(\ref{THREEB}) can be explicitly determined in all relevant situations where the continuity of the background fields is enforced. Furthermore, since the mode functions must always obey the Wronskian normalization condition of Eq. (\ref{TWOA3}) the determinant of the transition matrix 
must be equal to $1$
\begin{equation}
{\mathcal C}_{f\,f} \, {\mathcal C}_{g\,g} - {\mathcal C}_{f\,g}\,{\mathcal C}_{g\,f} =1.
\label{THREEC}
\end{equation}
By construction the actual expressions of the coefficients ${\mathcal C}_{f\,g}(k,\tau,\tau_{1})$ and ${\mathcal C}_{g\,f}(k,\tau,\tau_{1})$ vanish in the limit $\tau \to - \tau_{1}$ so that 
\begin{eqnarray}
&& \lim_{\tau \to - \tau_{1}} {\mathcal C}_{f\,f}(k,\tau,\tau_{1}) = \lim_{\tau \to -\tau_{1}} {\mathcal C}_{g\,g}(k,\tau,\tau_{1}) =1, 
\nonumber\\
&& \lim_{\tau \to -\tau_{1}} {\mathcal C}_{f\,g}(k,\tau,\tau_{1}) = \lim_{\tau \to -\tau_{1}} {\mathcal C}_{g\,f}(k,\tau,\tau_{1}) =0.
\label{THREEC2}
\end{eqnarray}
The entries of the transition matrix depend on the conformal time coordinate $\tau$,
on the transition time-scale $\tau_{1}$ and on the comoving wavenumber $k$.  In our 
specific situation, given the form of  Eqs. (\ref{THREEA})--(\ref{THREEB}), 
it is convenient to introduce the dimensionless wavenumber $x_{1} = k \, \tau_{1}$ and the shifted time variable $y \equiv y(\tau)= \tau + \tau_{1}(1 + q)$ with the notable property that $y(-\tau_{1}) = q\,\tau_{1}$; we finally recall that $ q= q(\epsilon, \delta)= (1- \epsilon) \delta$ has already been introduced  in Eq. (\ref{TWOC}). We then have, by definition 
that the arguments of the entries of the transition matrix do not separately 
depend on $k$ , $\tau_{1}$ and $\tau$ but just on $x_{1}$ and $k\, y$;
so, for instance, ${\mathcal C}_{f\,f}(k,\tau,\tau_{1}) = {\mathcal C}_{f\,f}(x_{1}, k\, y)$ and similarly for all the other entries of the transition matrix. For technical reasons it is finally practical to present a separate treatment of the cases $\delta > 1/2$,  $\delta \to 1/2$ and  $0< \delta < 1/2$.

\subsection{Explicit form of the spectral energy density for  $\delta > 1/2$}
When $\delta > 1/2$ the Universe can expand either faster (i.e. $\delta > 1$) or slower 
(i.e. $1/2 < \delta < 1$) than radiation but in both regimes the elements of the transition matrix consist products of Bessel functions with indices $\nu$ and $\nu+1$ where, in this 
case,  $\nu = \delta -1/2 >0$ \cite{abr1,abr2}. Their explicit form turns out to be:
\begin{eqnarray}
&& {\mathcal C}_{f\, f}(x_{1}, k\,y) = \pi \sqrt{q x_{1}/2} \sqrt{k\,y/2} \biggl[ J_{\nu+1}( q x_{1}) Y_{\nu}(k y) - Y_{\nu+1}(q x_{1}) J_{\nu}(k y) \biggr],
\nonumber\\
&& {\mathcal C}_{f\, g}(x_{1}, k\,y) = \pi \sqrt{q x_{1}/2} \sqrt{k\,y/2} \biggl[ J_{\nu}( q x_{1}) Y_{\nu}(k y) - Y_{\nu}(q x_{1}) J_{\nu}(k y) \biggr],
\nonumber\\
&& {\mathcal C}_{g\, f}(x_{1}, k\,y) = \pi \sqrt{q x_{1}/2} \sqrt{k\,y/2} \biggl[ Y_{\nu+1}( q x_{1}) J_{\nu +1}(k y) - J_{\nu+1}(q x_{1}) Y_{\nu+1}(k y) \biggr],
\nonumber\\
&& {\mathcal C}_{g\, g}(x_{1}, k\,y) = \pi \sqrt{q x_{1}/2} \sqrt{k\,y/2} \biggl[ Y_{\nu}( q x_{1}) J_{\nu+1}(k y) - Y_{\nu+1}(k y) J_{\nu}(q x_1) \biggr].
\label{THREED}
\end{eqnarray}
 As expected from Eq. (\ref{THREEC2}), for $\tau \to -\tau_{1}$ we have 
 that\footnote{This result  and
result follows from Eq. (\ref{THREEC2}) because of the explicit form of the Wronskian of Bessel functions \cite{abr1,abr2}. For $\tau \to - \tau_{1}$ we also have that $k\,y(- \tau_{1}) = q\, x_{1}$
so that, in this limit, all the arguments coincide.} ${\mathcal C}_{f\,f}(x_{1}, q x_{1}) = {\mathcal C}_{g\, g}(x_{1}, q x_{1}) =1$ ; in the same limit, ${\mathcal C}_{f\, g}(x_{1}, q x_{1}) = {\mathcal C}_{g\, f}(x_{1}, q x_{1}) =0$. If the post-inflationary evolution is dominated by a perfect irrotational fluid obeying all the energy conditions the maximal value of the barotropic index $w$ corresponds to the minimal expansion rate, i.e.  $\delta_{min} = 2/(3 w_{max} +1)$ and since $w_{max}$ can be at most $1$,  $\delta_{min} \to 1/2$. The post-inflationary expansion rate may also correspond to a stage dominated by an oscillating scalar field \cite{MAX2}; in this case we can always approximate the potential as $(\varphi/\overline{M_{P}})^{2 q}$ near the origin \cite{turn1} (see also \cite{turn2,turn3,turn4}). In this situation the coherent oscillations lead to $\delta = (q+1)/(2 q-1)$. For a quartic potential (i.e. $q \to 2$) we recover the case 
$\delta \to 1$ typical of a radiation phase while for $q \leq 2$ $\delta_{max} = 2$. The asymptote $\delta \to 1/2$ corresponds either to  $q \gg 1$ or to the absence of the potential. 

The spectral slopes and the amplitude of $\Omega_{gw}(k,\tau)$ will now be analyzed 
in all the relevant physical limits. From Eq. (\ref{TWOA9}) the expression of the spectral energy density can first be expressed through the elements of the transition matrix\footnote{The arguments of all the various functions appearing in Eq. (\ref{FIVEA}) have been omitted for the sake of conciseness but they will be reintroduced whenever needed.} given in Eq. (\ref{THREED})
\begin{equation}
\Omega_{gw}(k,\tau) = \frac{k^{5}}{6\, H^2\, \overline{M}_{P}^2 \, \pi^2 \, a^{4}} \biggl[
\biggl|{\mathcal C}_{f\, f} \, \overline{f}_{k} +  {\mathcal C}_{f\, g} \, \overline{g}_{k}/k\biggr|^2 
+ \biggl|{\mathcal C}_{g\, f} \,\overline{f}_{k} +  {\mathcal C}_{g\, g} \, \overline{g}_{k}/k\biggr|^2 \biggr],
\label{FIVEA}
\end{equation}
where, as already reminded in section \ref{sec1}, $\overline{M}_{P} = 1/\ell_{P} = M_{P}/\sqrt{8 \, \pi}$.  The two square moduli appearing in Eq. (\ref{FIVEA}) equally contribute to the final result but the first term appearing in each square modulus dominates against the second. To clarify this hierarchy we can formally rewrite Eq. (\ref{FIVEA}) as:
\begin{equation}
\Omega_{gw}(k,\tau) = \frac{k^{5}\, \bigl|\overline{f}_{k}(x_{1})\bigr|^2 }{6\, H^2\, \overline{M}_{P}^2 \, \pi^2 \, a^{4}} \biggl[ {\mathcal C}_{f\, f}^2  
\biggl| 1+  \biggl(\frac{{\mathcal C}_{f\, g} }{{\mathcal C}_{f\,f}}\biggr)\, \biggl(\frac{\overline{g}_{k}}{k\,\overline{f}_{k}}\biggr)\biggr|^2 
+ {\mathcal C}_{g\, f}^2 \biggl| 1 +  \biggl(\frac{{\mathcal C}_{g\, g}}{{\mathcal C}_{g\,f}}\biggr) \, \biggl(\frac{\overline{g}_{k}}{k\,\overline{f}_{k}}\biggr)\biggr|^2 \biggr],
\label{FIVEB}
\end{equation}
where we used that the entries of the transition matrix are real while 
$\overline{f}_{k}(x_{1})$ and $\overline{g}_{k}(x_{1})$ are both complex.
From Eqs. (\ref{THREEA}) and (\ref{THREEB})  $\overline{f}_{k}(x_{1}) = f_{k}(-\tau_{1})$ and $\overline{g}_{k}(x_{1})= g_{k}(-\tau_{1})$ and their ratio only depends in $x_{1}$ which is always smaller than $1$ as long as the frequencies are smaller than $\nu_{max}$. For $x_{1} > 1$ (i.e. 
$\nu > \nu_{max}$) the spectral energy density and the averaged multiplicity are exponentially suppressed \cite{BIRREL,PTOMS,FORDH,FORDL,mg0} (see also \cite{MAX1,MAX2} and the forthcoming discussion of section \ref{sec5}). This means that, as long as $x_{1} < 1$ (i.e. $k < a_{1} \, H_{1}$), the momentum mode function $\overline{g}_{k}$ is systematically smaller than $k \, \overline{f}_{k}$
\begin{equation}
\frac{\overline{g}_{k}(x_{1})}{k\,\,\overline{f}_{k}(x_{1})} = - \frac{H_{\mu-1}^{(1)}(x_{1})}{H_{\mu}^{(1)}(x_{1})} = 
- \frac{(1 - \epsilon) \, x_{1}}{(1 + \epsilon)} = - \frac{q(\epsilon,\delta)}{\delta} x_{1} [ 1 - \epsilon + {\mathcal O}(\epsilon^2)],
\label{FIVEC}
\end{equation}
while for $x_{1} > 1$ the spectral energy density is suppressed because of the unitarity 
of the process of graviton production. Therefore, thanks to Eq. (\ref{FIVEC}), both expressions appearing in the square moduli of Eq. (\ref{FIVEB}) are ${\mathcal O}(x_{1}^2)$ and therefore subleading when $x_{1} < 1$ (i.e. $\nu< \nu_{max}$, see Eq. (\ref{kx1}) and discussion therein). From a quantitative viewpoint in the case $\delta >1/2$ we would have 
\begin{equation}
\lim_{x_{1} < 1} x_{1} \, \frac{{\mathcal C}_{f\,g}(x_{1}, k\,y)}{{\mathcal C}_{f\,f}(x_{1}, k\,y)} =
\lim_{x_{1} < 1} x_{1} \, \frac{{\mathcal C}_{g\,g}(x_{1}, k\,y)}{{\mathcal C}_{g\,f}(x_{1}, k\,y)} = 
\frac{(1 -\epsilon) \, \delta\, x_{1}^2}{(2 \delta -1)} \ll 1.
\label{FIVEEaa}
\end{equation}
Equation (\ref{FIVEEaa}) concretely demonstrates that the expansion scheme is not well defined for  $\delta \to 1/2$ 
since the obtained results, in this limit, are formally divergent. This is why for
$\delta\to 1/2$ there are logarithmic corrections that may affect the final expression 
of the spectral energy density; these issues are separately analyzed in the following subsection\footnote{For the moment we just remark that, provided $\delta > 1/2$ the contributions appearing in Eq. (\ref{FIVEB}) can be analytically estimated
with the negligible correction given in Eq. (\ref{FIVEEaa}).}.  
 
 The results of Eqs. (\ref{FIVEC}) and (\ref{FIVEEaa}) do not involve any assumption on $k \, y(\tau)$ and they hold for all the amplified wavelengths (i.e. $k < a_{1} H_{1}$ and $\nu < \nu_{max}$) without 
further restrictions. The physical result we are after should however involve all the 
wavelengths that already reentered the Hubble radius (i.e. $k y(\tau) \simeq k \tau \gg 1$).
 Since the leading contribution to the spectral energy density of Eq. (\ref{FIVEB}) follows from ${\mathcal C}_{f\, f}( x_{1}, k\,y)$ and ${\mathcal C}_{f\, g}(x_{1}, k\,y)$ we first write these entries in the limit $x_{1} <1$:
\begin{eqnarray}
{\mathcal C}_{f\, f}(x_{1}, k\,y) &=& \bigl(q x_{1}/2 \bigr)^{-\delta}  \sqrt{k\,y/2} \, J_{\delta-1/2}(k\, y) \Gamma(\delta +1/2)[1  + {\mathcal O}(x_{1})],
\nonumber\\
{\mathcal C}_{g\, f}(x_{1}, k\,y) &=& - \bigl(q x_{1}/2\bigr)^{-\delta} \sqrt{k\,y/2} \, J_{\delta+1/2}(k\, y) \Gamma(\delta +1/2)[1 + {\mathcal O}(x_{1})]. 
\label{SIXEEb}
\end{eqnarray}
When the results of Eq. (\ref{SIXEEb}) are combined with Eq. (\ref{FIVEB}) the rapid oscillations  
get suppressed in the limit $k\, y > 1$ since  $k \, y[J^2_{\delta -1/2}(k\, y) + J^2_{\delta +1/2}(k\, y)] \to 2/\pi$ as it follows from the large-argument limit of the Bessel functions \cite{abr1,abr2}. It is finally important 
to appreciate that the limits $x_{1} < 1$ and $k\, y >1$ commute and can be simultaneously enforced
\begin{eqnarray}
&&\biggl| {\mathcal C}_{f\, f}(x_{1}, k\,y) \, - \frac{H^{(1)}_{\mu-1}(x_{1})}{H_{\mu}^{(1)}(x_{1})} {\mathcal C}_{f\, g}(x_{1}, k\,y)\biggr|^2 
+ \biggl| {\mathcal C}_{g\, f}(x_{1}, k\,y)  - \frac{H^{(1)}_{\mu-1}(x_{1})}{H_{\mu}^{(1)}(x_{1})} {\mathcal C}_{g\, g}(x_{1}, k\,y) \biggr|^2 
\nonumber\\
&&= \frac{1}{\pi} \,\bigl(q x_{1}/2\bigr)^{-2 \delta}\,\biggl[ \Gamma^2(\delta+1/2) + {\mathcal O}(x_{1}^2) + {\mathcal O}( |k \, y|^{-2})\biggr].
\label{SIXEEbb}
\end{eqnarray}
Either from Eqs. (\ref{FIVEC})--(\ref{FIVEEaa}) or from Eqs. (\ref{SIXEEb})--(\ref{SIXEEbb}) we can then deduce the spectral energy density of Eqs. (\ref{FIVEA})--(\ref{FIVEB}) and obtain
\begin{equation}
\Omega_{gw}(k,\tau) = \overline{\Omega}(r_{T}, \delta, H_{1}, H_{r}) \biggl(\frac{k}{a_{1}\, H_{1}}\biggr)^{ n_{T}(\delta,r_{T})}\biggl[ 1 + {\mathcal O}\biggl(\frac{k^2}{a_{1}^2 \, H_{1}^2}\biggr) \biggr].
\label{SIXEEc}
\end{equation}
Equation (\ref{SIXEEc}) is valid when all the wavelengths of the spectrum are shorter than the Hubble radius; 
furthermore the value of $n_{T}(\delta, r_{T})$ coincides exactly
with the WKB result already obtained in Eq. (\ref{SPIN1}). All the numerical factors entering $\Omega_{gw}(k,\tau)$ have been included inside the overall normalization $\overline{\Omega}(r_{T}, \delta, H_{1}, H_{r})$
\begin{equation}
\overline{\Omega}(r_{T}, \delta, H_{1}, H_{r}) ={\mathcal B}(\delta, r_{T}) \,r_{T}\, {\mathcal A}_{{\mathcal R}} \, \Omega_{R\,0} 
\biggl(\frac{g_{\rho,\, r}}{g_{\rho,\,eq}}\biggr) \biggl(\frac{g_{s,\,eq}}{g_{s,\,r}}\biggr)^{4/3} 
\biggl(\frac{H_{r}}{H_{1}}\biggr)^{2 (\delta-1)/(\delta+1)}.
\label{SPEN1}
\end{equation}
The meanings of $H_{1}$ and $H_{r}$ appearing in Eq. (\ref{SPEN1}) coincide with the ones already introduced in the WKB analysis (see Eqs. (\ref{OMWK1})--(\ref{OMWK2}) and discussions therein); as before the value of $(H_{1}/M_{P})$ follows from the amplitude to the scalar fluctuations of the geometry. Finally, in Eq. (\ref{SPEN1}) the function ${\mathcal B}(\delta, r_{T})$ is defined as:
\begin{equation}
{\mathcal B}(\delta, r_{T})=\frac{2^{2(\mu + \delta) -3}}{3 \, \pi^2\, q^{2 \delta}}\, \Gamma^2(\mu)\, \Gamma^2(\delta +1/2), \qquad 
\mu = \frac{ 48 - r_{T}}{32 -2\,r_{T}},
\label{SPEN2}
\end{equation}
has been also introduced for convenience and it now contains the numerical factors that are essential for a comparison 
between the WKB approach and the present strategy. The largest 
value of ${\mathcal B}(\delta, r_{T})$ corresponds to the limit $r_{T} \to 0$ where ${\mathcal B}(\delta,0) = q^{- 2 \delta} \, \Gamma^2(\delta +1/2)/(12 \pi)$. In the limit $\delta \to 1$ the dependence on $H_{r}$ fully disappear from Eq. (\ref{SPEN1}) while $n_{T}(1, r_{T}) \to - r_{T}/8 + {\mathcal O}(r_{T}^2)$ which is the value of quasi-flat spectrum of the tensor modes when the consistency relations are enforced.

\subsection{The case $\delta \to 1/2$ and its spectral features}
As already mentioned in Eq. (\ref{FIVEEaa}) the concurrent  
expansions in terms of $x_{1}$ and $k\, y$ are not well defined when $\delta \to 1/2$. In this 
physical case the expansion rate is slower than radiation and the entries of the transition matrix are: 
 \begin{eqnarray}
 && {\mathcal C}_{f\, f}(x_{1}, k\, y) = \pi \sqrt{q x_{1}/2} \sqrt{ k\,y/2} \biggl[ J_{1}( q x_{1}) Y_{0}(k y) - Y_{1}(q x_{1}) J_{0}(k y) \biggr],
\nonumber\\
&& {\mathcal C}_{f\, g}( x_{1}, k\,y) = \pi \sqrt{q x_{1}/2} \sqrt{ k\,y/2} \biggl[ J_{0}( q x_{1}) Y_{0}(k y) - Y_{0}(q x_{1}) J_{0}(k y) \biggr],
\nonumber\\
&& {\mathcal C}_{g\, f}(x_{1}, k\,y) = \pi \sqrt{q x_{1}/2} \sqrt{ k\, y/2} \biggl[ Y_{1}( q x_{1}) J_{1}(k y) - J_{1}(q x_{1}) Y_{1}(k y) \biggr],
\nonumber\\
&& {\mathcal C}_{g\, g}(x_{1}, k\,y) = \pi\sqrt{q x_{1}/2} \sqrt{ k\,y/2} \biggl[ Y_{0}( q x_{1}) J_{1}(k y) - Y_{1}(k y) J_{0}(q x_1) \biggr].
\label{THREED0}
\end{eqnarray}
The results of Eq. (\ref{THREED0}) may also follows from Eq. (\ref{FOURB}) by recalling that, in general terms, $Y_{-1}(z) = - Y_{1}(z)$ and $J_{-1}(z) = - J_{1}(z)$. We know that the small argument limit of $Y_{0}(z)$ is logarithmically divergent \cite{abr1,abr2}; as suggested long ago and this is the reason why a series of logarithmic 
corrections must be included in the expression of the transfer function \cite{mg0}. 
This is why Eq. (\ref{THREED0}) already suggests the reason of the logarithmic divergences potentially appearing in the expansion of Eq. (\ref{FIVEEaa}) for $\delta \to 1/2$.
In the limit $\delta \to 1/2$ we have that $q(\delta,\epsilon) = \delta(1- \epsilon) \simeq 1/2$ and if we expand the spectral energy density for $q\, x_{1} \ll 1$  the 
partial results for the two contributions appearing in the spectral energy density become:
\begin{eqnarray}
{\mathcal C}_{f\,f}(x_{1},k\,y) -  x_{1} {\mathcal C}_{f\,g}(x_{1}, k\,y) &=& 
\frac{k y}{q x_{1}} J_{0}^2(k\,y) + (k y/2) J_{0}(k\,y)\biggl\{J_{0}(k\,y) \biggl[1+ 6 \gamma + 6 \ln{(q x_{1}/2)}\biggr] 
\nonumber\\
&-& 3 \pi Y_{0}(k\,y)  \biggr\} \,q x_{1} + {\mathcal O}(q^3 x_{1}^3),
\nonumber\\
 {\mathcal C}_{g\,f}(x_{1},k\,y) -  x_{1} {\mathcal C}_{g\,g}(x_{1}, k\,y) &=& 
\frac{k y}{q x_{1}} J_{1}^2(k\,y) + (k y/2) J_{1}(k\,y)\biggl\{J_{1}(k\,y) \biggl[1+ 6 \gamma + 6 \ln{(q\,x_{1}/2)}\biggr]
\nonumber\\
&-& 3 \pi Y_{1}(k\,y)  \biggr\} \,q x_{1} + {\mathcal O}(q^3 x_{1}^3).
\label{FIRST0}
\end{eqnarray}
where $\gamma=0.5772$ is the Euler-Mascheroni constant. We can then 
compute explicitly the contribution to the spectral energy density 
\begin{eqnarray}
&& \biggl|{\mathcal C}_{f\,f}(x_{1},\,k y) -  x_{1} {\mathcal C}_{f\,g}(x_{1}, k y)\biggr|^2 + \biggl|{\mathcal C}_{g\,f}(k,\,x_{1},\,k y) -  x_{1} {\mathcal C}_{g\,g}(k, x_{1}, k y)\biggr|^2 =
\nonumber\\
&& \frac{k \,y}{q \, x_{1}} \biggl[ J_{0}^2(k\,y) + J_{1}^2(k\, y)\biggr]  
 +  q x_{1} \bigl(k\, y/2\bigr)\biggl\{ - 3 \pi \biggl[J_{0}(k\, y) Y_{0}(k\, y) 
+ J_{1}(k\, y) Y_{1}(k\, y)\biggr] 
\nonumber\\
&& + \biggl[ J_{0}^2(k\,y) + J_{1}^2(k\, y)\biggr]\biggl[ 1 + 6 \gamma - 6 \ln{(q\, x_{1}/2)}\biggr]\biggr\}.
\label{SECOND0}
\end{eqnarray}
Again the terms containing the combinations of the type $J_{0}^2(z)$ and of $J_{1}^2(z)$ have a well defined large-argument limit which is of the order of $ 2/(\pi \, z)$; this is also the case 
for the combination $Y_{0}^2(z)+ Y_{1}^2(z) \to 2/(\pi\, z)$ for $z\gg 1$. The mixed combinations of the type $J_{0}(z) Y_{0}(z) + J_{1}(z) Y_{1}(z)$ oscillate around $0$ in the limit $z \gg 1$ and are subleading the perturbative expansion. For this reason the right-hand side of Eq. (\ref{SECOND0}) simply becomes:
\begin{equation}
\frac{2}{\pi q \, x_{1}} +  \frac{q x_{1}}{\pi} \biggl[ 1 + 6 \gamma - 6 \ln{(q\, x_{1}/2)}\biggr].
\label{THIRD0}
\end{equation}
The same strategy leading to Eq. (\ref{THIRD0}) can be enforced order by order; for 
instance the following correction to Eq. (\ref{THIRD0}) reads:
\begin{eqnarray}
 \frac{q^3 x_{1}^3}{16 \pi} &\biggl\{&18 \pi^2
 + 29 + 4 \ln{2}  +72 (\gamma^2 + \ln^2{2}) - 4 \gamma ( 1 + 36 \ln{2}) 
 \nonumber\\
&+& 4 \ln{(q x_{1})}\biggl[ - 1 + 36 \gamma + 18 \ln{(q x_{1})} - 36 \ln{2}\biggr]\biggr\}.
\label{FOURTH0}
\end{eqnarray}
As a final step we can therefore deduce the expression of the spectral energy density 
valid for $\delta \to 1/2$ and in the limit where all the wavelengths of the spectrum are shorter than the Hubble radius:
\begin{equation}
\Omega_{gw}(k,\tau) = \frac{4}{3 \pi^2 q} \biggl(\frac{H_{1}}{M_{P}}\biggr)^{2} \biggl(\frac{H_{1}^2 \, a_{1}^4}{H^2 \, a^4} \biggr) \biggl(\frac{k}{a_{1}\, H_{1}}\biggr)^{m_{T}(r_{T})} \biggl[ 1 + q^2 x_{1}^2 \ell_{1}(x_{1}) + q^3 x_{1}^3 \ell_{2}(x_{1}) + {\mathcal O}(q^5 x_{1}^5)\biggr],
\label{FIFTH0}
\end{equation}
where the higher order terms of the expansion also contain the same kind of logarithmic corrections 
appearing in  the two functions $\ell_{1}(x_{1})$ and $\ell_{2}(x_{1})$ 
\begin{eqnarray}
\ell_{1}(x_{1}) &=& \frac{1}{2} \biggl[ 1 + 6 \gamma + 6 \ln{\bigl(q x_{1}/2\bigr)}\biggr],
\nonumber\\
\ell_{2}(x_{1}) &=&\frac{1}{32} \biggl[ 18 \pi^2 + 4 \ln{2} + \bigl(72 \gamma^2 - 4 \gamma + 29\bigr)
 \nonumber\\
 &+& 144 (\gamma - \ln{2}-1) \ln{\bigl(q x_{1}/2\bigr)}
 + 72 \,\ln^2{\bigl(q x_{1}/2\bigr)} \biggr].
 \label{SIXTH0}
 \end{eqnarray}
In Eq. (\ref{FIFTH0}) the spectral index $m_{T}(r_{T})$ coincides with $n_{T}(\delta,r_{T})$ 
introduced in Eq. (\ref{SIXEEc}) in the limit $\delta \to 1/2$:
\begin{equation}
m_{T}(r_{T}) = \lim_{\delta \to 1/2} \, n_{T}(\delta, r_{T}) = \frac{16 - 3 \, r_{T}}{16 - r_{T}}.
\label{FIFTH0a}
\end{equation}

\subsection{The transition matrix for $0 < \delta < 1/2$}
The last case we are going to discuss involves the range $0 < \delta < 1/2$; also in this 
situation the expansion rate is always slower than radiation and the explicit form of 
the transition matrix is:
\begin{eqnarray}
{\mathcal C}_{f\, f}(x_{1}, k\,y) &=& \pi \sqrt{q x_{1}/2} \sqrt{ k\,y/2} \biggl[ Y_{\nu-1}( q x_{1}) J_{\nu}(k y) - J_{\nu-1}(q x_{1}) Y_{\nu}(k y) \biggr],
\nonumber\\
{\mathcal C}_{f\, g}(x_{1}, k\,y) &=& \pi \sqrt{q x_{1}/2} \sqrt{ k\,y/2} \biggl[ J_{\nu}( q x_{1}) Y_{\nu}(k y) - Y_{\nu}(q x_{1}) J_{\nu}(k y) \biggr],
\nonumber\\
{\mathcal C}_{g\, f}(x_{1}, k\, y) &=& \pi \sqrt{q x_{1}/2} \sqrt{ k\,y/2} \biggl[ Y_{\nu-1}( q x_{1}) J_{\nu -1}(k y) - J_{\nu-1}(q x_{1}) Y_{\nu-1}(k y) \biggr],
\nonumber\\
{\mathcal C}_{g\, g}(x_{1}, k\,y) &=& \pi \sqrt{q x_{1}/2} \sqrt{ k\,y/2} \biggl[ J_{\nu}( q x_{1}) Y_{\nu-1}(k y) - J_{\nu-1}(k y) Y_{\nu}(q x_1) \biggr],
\label{FOURB}
\end{eqnarray}
with the further difference that in Eq. (\ref{FOURB}) $\nu= 1/2 - \delta >0$.
In spite of the differences between Eqs. (\ref{FOURB}) and (\ref{THREED}) 
the spectral energy density in critical units ultimately coincide in the two cases. To 
verify this result we can use the conclusions of the previous analysis implying 
that two physical limits $x_{1} < 1$ and $k\, y> 1$ commute; we can therefore 
expand the result of Eq. (\ref{FOURB}) and the result is:
\begin{eqnarray}
{\mathcal C}_{f\,f}(x_{1}, y) &=& - \bigl(q x_{1}/2\bigr)^{-\delta} \biggl\{\frac{\Gamma(\delta+1/2)}{\sqrt{\pi}}\, \cos{[\pi (\delta +1/2)]} \cos{[ k y + \pi (\delta -1)/2]} 
\nonumber\\
&+& \frac{\sqrt{\pi}}{\Gamma(1/2 - \delta)}\sin{[k y + \pi(\delta -1)/2]} + {\mathcal O}(x_{1}) + {\mathcal O}(|k\,y|^{-2})\biggr\},
\nonumber\\
{\mathcal C}_{f\,g}(k, x_{1}, y) &=& \bigl(q x_{1}/2 \bigr)^{\delta}\biggl\{ \frac{\Gamma(1/2-\delta)}{\sqrt{\pi}} \cos{[ k y + \pi (\delta -1)/2]} + {\mathcal O}(x_{1}) + {\mathcal O}(|k\,y|^{-2})\biggr\},
\nonumber\\
{\mathcal C}_{g\,f}(k, x_{1}, y) &=& - \bigl(q x_{1}/2 \bigr)^{-\delta} \biggl\{\frac{\Gamma(\delta+1/2)}{\sqrt{\pi}}\, \cos{[\pi(\delta+1/2)]} \cos{[ k y + \pi\delta/2]} 
\nonumber\\
&+& \frac{\sqrt{\pi}}{\Gamma(1/2 - \delta)}\sin{[k y + \pi\delta/2]} + {\mathcal O}(x_{1}) + {\mathcal O}(|k\,y|^{-2})\biggr\},
\nonumber\\
{\mathcal C}_{g\,g}(k, x_{1}, y) &=& \bigl(q x_{1}/2 \bigr)^{\delta}\biggl\{ \frac{\Gamma(1/2-\delta)}{\sqrt{\pi}} \cos{[ k y + \pi \delta/2]} + {\mathcal O}(x_{1}) +{ \mathcal O}(|k\,y|^{-2})\biggr\}.
\label{FOURBa}
\end{eqnarray}
From Eq. (\ref{FOURBa}) we can verify the validity of all the approximations 
employed so far. This means that to leading order in $(q x_{1})$ and $|k \, y|^{-2}$ the spectral 
energy in critical units can also be written as 
\begin{equation}
\Omega_{gw}(k,\tau) = \frac{k^{5}\, \bigl|\overline{f}_{k}(x_{1})\bigr|^2 }{6\, H^2\, \overline{M}_{P}^2 \, \pi^2 \, a^{4}} \biggl[ {\mathcal C}_{f\, f}^2(k, x_{1},y) \,+  {\mathcal C}_{g\, f}^2(k, x_{1},y) + {\mathcal O}(x_{1}^2)\biggr].
\label{FIVED}
\end{equation}
If Eq. (\ref{FOURBa}) is now inserted into Eq. (\ref{FIVED}) the spectral energy density in critical units becomes:
\begin{equation}
\Omega_{gw}(k,\tau) = \overline{\Omega}(r_{T}, \delta, H_{1}, H_{r}) \biggl(\frac{k}{a_{1}\, H_{1}}\biggr)^{ n_{T}(\delta,r_{T})}
\biggl[ 1 + {\mathcal O}\biggl(\frac{k^2}{a_{1}^2 \, H_{1}^2}\biggr) \biggr], 
\label{SIXEEca}
\end{equation}
which coincides exactly with the result of Eq. (\ref{SIXEEc}) except for the explicit 
form of ${\mathcal B}(\delta, r_{T})$ that now reads:
\begin{equation}
{\mathcal B}(\delta, r_{T})=\frac{2^{2\mu -3}}{3 \, \pi^2\, q^{2 \delta}}\, \Gamma^2(\mu)\, \Gamma^2(\delta +1/2) \biggl\{ \cos^2{[\pi (\delta +1/2)]} + \frac{\pi^2}{\Gamma^2(1/2- \delta) \, \Gamma^2(1/2 + \delta)} \biggr\},
\label{SPEN2a}
\end{equation}
where, as before, $\mu = (48 - r_{T})/(16 - r_{T})$. While Eqs. (\ref{SPEN2}) and (\ref{SPEN2a}) 
seem superficially different but they lead to the same condition\footnote{Because 
of the usual identities involving the Gamma functions we have that $\pi \sin{\pi \delta} = \Gamma(1/2- \delta) \, \Gamma(1/2 + \delta)$; this observation implies immediately that the quantity within the squared brackets is just $1$.}. Although the results of Eqs. (\ref{SIXEEc}) and (\ref{FIVED}) are formally the same they apply for two different ranges of $\delta$, i.e. $\delta >1/2$ and $0\leq \delta <1/2$ respectively.

\subsection{Comparison of the different strategies}
The spectral energy densities in critical units determined from the WKB approximation 
and from the transition matrix are overall consistent  and the mismatch between the two complementary 
approaches can be unambiguously quantified. This analysis is useful since 
the  analytic determinations can be reasonably employed in order  
 to normalize the high frequency branch of the spectrum.
Let us first consider the case where $\delta \neq 1/2$ and denote 
by ${\mathcal D}(\delta, r_{T})$ the ratio between the 
WKB estimate (see Eqs. (\ref{OMWK2a})  and (\ref{OMWK2c1}))and the result based on the transition matrix (see Eqs. (\ref{SIXEEc}) and (\ref{SIXEEca})):
\begin{equation}
{\mathcal D}(\delta, r_{T})  = \frac{c_{1}(r_{T}/16)}{12 \, {\mathcal B}(\delta, r_{T})\, q^{ 2 \delta}}.
\label{COMP1}
\end{equation}
With simple algebra involving the explicit form of ${\mathcal B}(\delta, r_{T})$ (see, for instance, Eq. (\ref{SPEN2})) 
Eq. (\ref{COMP1}) becomes 
\begin{equation}
{\mathcal D}(\delta, r_{T}) = \frac{4 \pi^2 [ 1024 +(16 -r_{T})^2]}{2^{2 \delta + (48 -r_{T})/(16 -r_{T})} (48 - r_{T})^2 \,\Gamma^2(\delta +1/2) \, \Gamma^2[(48 - r_{T})/(32 - r_{T})]}.
\label{COMP1a}
\end{equation}
To appreciate the numerical relevance of the mismatch between the two approaches we can consider the limit of Eq. (\ref{COMP1a}) $r_{T} \to 0$ and
obtain:
\begin{equation}
\lim_{r_{T} \to 0} {\mathcal D}(\delta, r_{T})  = \frac{ 10\, \pi}{9} \, \frac{\, 2^{-2\delta}}{ \Gamma^2(\delta +1/2)}.
\label{COMP2}
\end{equation}
Equations (\ref{COMP1a}) and (\ref{COMP2}) show that the WKB approximation always underestimates the spectral energy density for very large 
values of $\delta$ but it is consistent with the exact result within a factor of $10$ for $\delta < 3$. In the case $\delta \to 1/2$ the logarithmic corrections obtained within the WKB approach follow from Eqs. (\ref{INT4}) and (\ref{WKB11a}). The comparison with Eqs. (\ref{FIFTH0})--(\ref{SIXTH0}) demonstrates that the structure of the logarithmic corrections is different within the two approaches. The approximations can be also compared with the results based on the exact form

\renewcommand{\theequation}{5.\arabic{equation}}
\setcounter{equation}{0}
\section{The accuracy of high frequency normalization}
\label{sec5}
The approximations discussed in the previous sections are mutually consistent and they can be used to assess various properties of the high frequency signal without preliminary knowledge of the spectrum at lower frequencies. While this procedure inevitably contains some inaccuracies that are quantified hereunder, the semi-analytic approach is often swifter than the full numerical analysis.  Within this logic the first point will now be to scrutinize (both exactly and approximately) the averaged multiplicity of the produced pairs and its exponential suppression. After clarifying the scaling properties of the averaged multiplicity 
the same analysis shall be extended to the spectral energy density. In the second part of this section we are going to compare the high and low frequency normalizations imposed, respectively, on  the analytic spectrum (that disregards the low frequency region) and on the numerical results (that take also into account the late-time sources of suppression in the nHz region). The bounds on the post-inflationary expansion history and the expected ranges of the chirp amplitudes are finally analyzed in the last part of the section. We are going to argue that for a direct detection of a cosmic signal in the ultra-high frequency band the required chirp amplitudes should be at least twelve orders of magnitude smaller than the ones currently measured in the audio band ranging between few Hz and ten kHz. 

\subsection{Scaling of the averaged multiplicity and the single-graviton limit}
The approximation schemes of sections \ref{sec3} and \ref{sec4}
are based on the enforcement of the two concurrent limits namely $x_{1} <1$ and $k\,y > 1$. While both requirements are physically justified the former is more general than the latter insofar as it is always accurately verified below the maximal frequency domain (i.e. $\nu < \nu_{max}$). The averaged multiplicity is the appropriate dimensionless variable that encodes the relevant information on the production of the pairs of gravitons and this is why it is practical to study it both exactly and approximately. In a quantum mechanical perspective, the amplification of the mode functions corresponds to the production of graviton pairs with opposite three-momenta. Indeed, from the explicit form of the quantum Hamiltonian we have \cite{MAX2}:
\begin{equation}
\widehat{H}_{g}(\tau) = \frac{1}{2} \int d^{3} k \sum_{\alpha=\oplus,\, \otimes} \biggl\{ k \,\,\biggl[ \widehat{a}^{\dagger}_{\vec{k},\,\alpha} \widehat{a}_{\vec{k},\,\alpha} 
+ \widehat{a}_{- \vec{k},\,\alpha} \widehat{a}^{\dagger}_{-\vec{k},\,\alpha} \biggr] 
+ \lambda \,\,\widehat{a}^{\dagger}_{-\vec{k},\,\alpha} \widehat{a}_{\vec{k},\,\alpha}^{\dagger} 
+ \lambda^{\ast} \,\,\widehat{a}_{\vec{k},\,\alpha} \widehat{a}_{-\vec{k},\,\alpha}  \biggr\},
\label{MULT0a}
\end{equation}
where  $ \lambda = i {\mathcal H} = i \, a \, H$. The three 
classes of terms quadratic in the creation and annihilation operators are in fact the generators of the $SU(1,1)$ group 
and the evolution equations for $\widehat{a}_{\vec{k}}$ and $\widehat{a}_{-\vec{k},\,\alpha}^{\dagger}$
in the Heisenberg description follow from the Hamiltonian (\ref{MULT0a}): 
\begin{eqnarray}
\frac{d \widehat{a}_{\vec{k},\,\alpha}}{d\tau} &=& i \, [ \widehat{H}_{g},\widehat{a}_{\vec{k},\,\alpha}] =  - i\, k \, \widehat{a}_{\vec{k},\,\alpha} -  i \, \lambda \widehat{a}_{- \vec{k},\,\alpha}^{\dagger},
\nonumber\\
\frac{d \widehat{a}_{-\vec{k},\,\alpha}^{\dagger}}{d\tau} &=& i \, [ \widehat{H}_{g},\widehat{a}^{\dagger}_{-\vec{k},\,\alpha}]=  i\, k \, \widehat{a}_{-\vec{k},\,\alpha}^{\dagger} +  i \, \lambda^{\ast} \widehat{a}_{\vec{k},\,\alpha}.
\label{MULT0b}
\end{eqnarray}
The explicit evolution of the creation and annihilation operators given by Eq. (\ref{MULT0b}) can be solved by introducing  
two  (complex) functions $u_{k\,\alpha}(\tau)$ and $v_{k,\,\alpha}(\tau)$:
\begin{eqnarray}
\widehat{a}_{\vec{k}, \, \alpha}(\tau) &=& u_{k,\,\alpha}(\tau) \, \widehat{b}_{\vec{k},\, \alpha} -  v_{k,\,\alpha}(\tau) \,  \widehat{b}_{-\vec{k},\, \alpha}^{\,\dagger}, 
\nonumber\\
\widehat{a}_{-\vec{k}, \, \alpha}^{\,\dagger}(\tau) &=& u_{k,\,\alpha}^{\ast}(\tau) \, \widehat{b}_{-\vec{k},\, \alpha}^{\,\dagger} -  v_{k,\,\alpha}^{\ast}(\tau) \,  \widehat{b}_{\vec{k},\, \alpha}.
\label{MULT1}
\end{eqnarray}
The complex functions $u_{k}(\tau)$ and $v_{k}(\tau)$ can be directly related to the mode functions defined earlier on 
$f_{k}(\tau) = [u_{k}(\tau) - v_{k}^{\ast}(\tau)]/\sqrt{ 2 k}$ and as $g_{k}(\tau) = - i \, \sqrt{k/2} [u_{k}(\tau) + v_{k}^{\ast}(\tau)]$. 
Because of the unitarity of the evolution the two complex functions are subjected to the constraint 
$|u_{k,\,\alpha}(\tau)|^2 - |v_{k,\,\alpha}(\tau)|^2 = 1$;  the averaged multiplicity of the produced gravitons with opposite three-momenta 
follows from Eq. (\ref{MULT1}) and it is given by
\begin{equation}
\langle \widehat{N}_{k}(\tau) \rangle = \sum_{\alpha = \oplus,\otimes} \biggl[\langle \widehat{N}_{\vec{k},\alpha}(\tau) \rangle + \langle \widehat{N}_{-\vec{k},\alpha}(\tau) \rangle\biggr] = 4 \,|v_{k}(\tau)|^2.
\label{MULT2}
\end{equation}
The factor $4$ counts the gravitons with opposite three-momenta summed over the two polarizations and this means that 
in Eq. (\ref{MULT2})  $\overline{n}(k,\tau) = |v_{k}(\tau)|^2$ denotes the multiplicity of the pairs associated with a single tensor polarization.  Thus the averaged multiplicity depends on the elements of the transition matrix and solely depends on the dimensionless variables $x_{1}$ and $k\,y$:
\begin{eqnarray}
\overline{n}(x_{1}, k\, y) &=& \frac{\pi \, x_{1}}{8} \biggl\{ \biggl[ {\mathcal C}_{f\,f}^2(x_{1}, k\, y) + {\mathcal C}_{g\,f}^2(x_{1}, k\,y) \biggr]
\biggl[ J_{\mu}^2(x_{1}) + Y_{\mu}^2(x_{1})\biggr] 
\nonumber\\
&+& \biggl[{\mathcal C}_{f\,g}^2(x_{1}, k\, y) + {\mathcal C}_{g\,g}^2(x_{1}, k\, y) \biggr]
\biggl[ J_{\mu-1}^2(x_{1}) + Y_{\mu-1}^2(x_{1})\biggr]
\nonumber\\
&-& 2 \biggl[ {\mathcal C}_{f\,f}(x_{1}, k\, y)\,{\mathcal C}_{f\,g}(x_{1}, k\, y) + {\mathcal C}_{g\,f}(x_{1}, k\, y)\,{\mathcal C}_{g\,g}(x_{1}, k\, y)\biggr]\biggr\}.
\label{MULT3}
\end{eqnarray}
We stress that the averaged multiplicity counts in fact the produced gravitons in spite of the vacuum contribution 
which can be always renormalized \cite{EXP1,MAX1,MAX2} but this is mmaterial when the averaged multiplicities are very large.
 \begin{figure}[!ht]
\centering
\includegraphics[height=5.7cm]{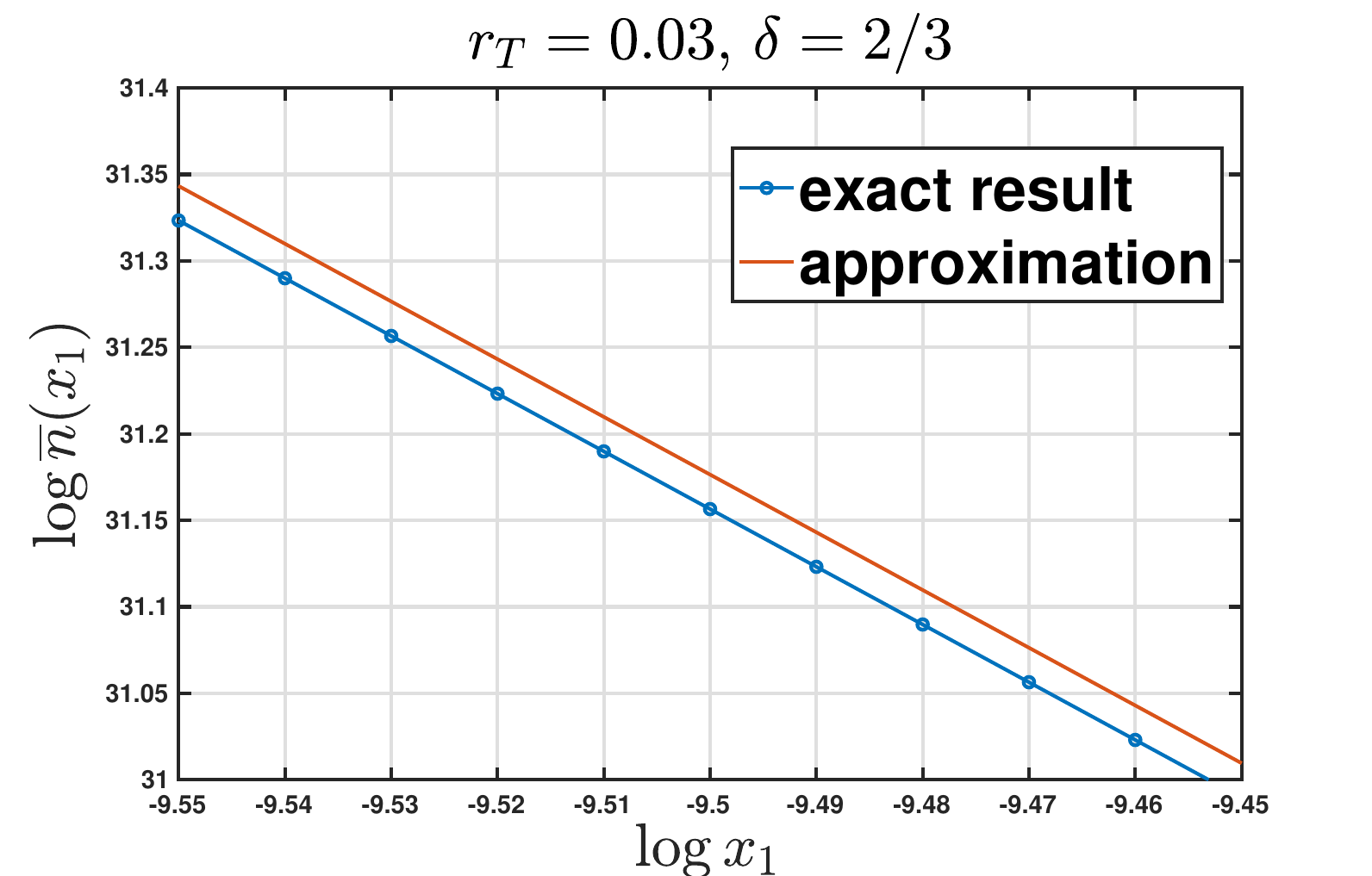}
\includegraphics[height=5.7cm]{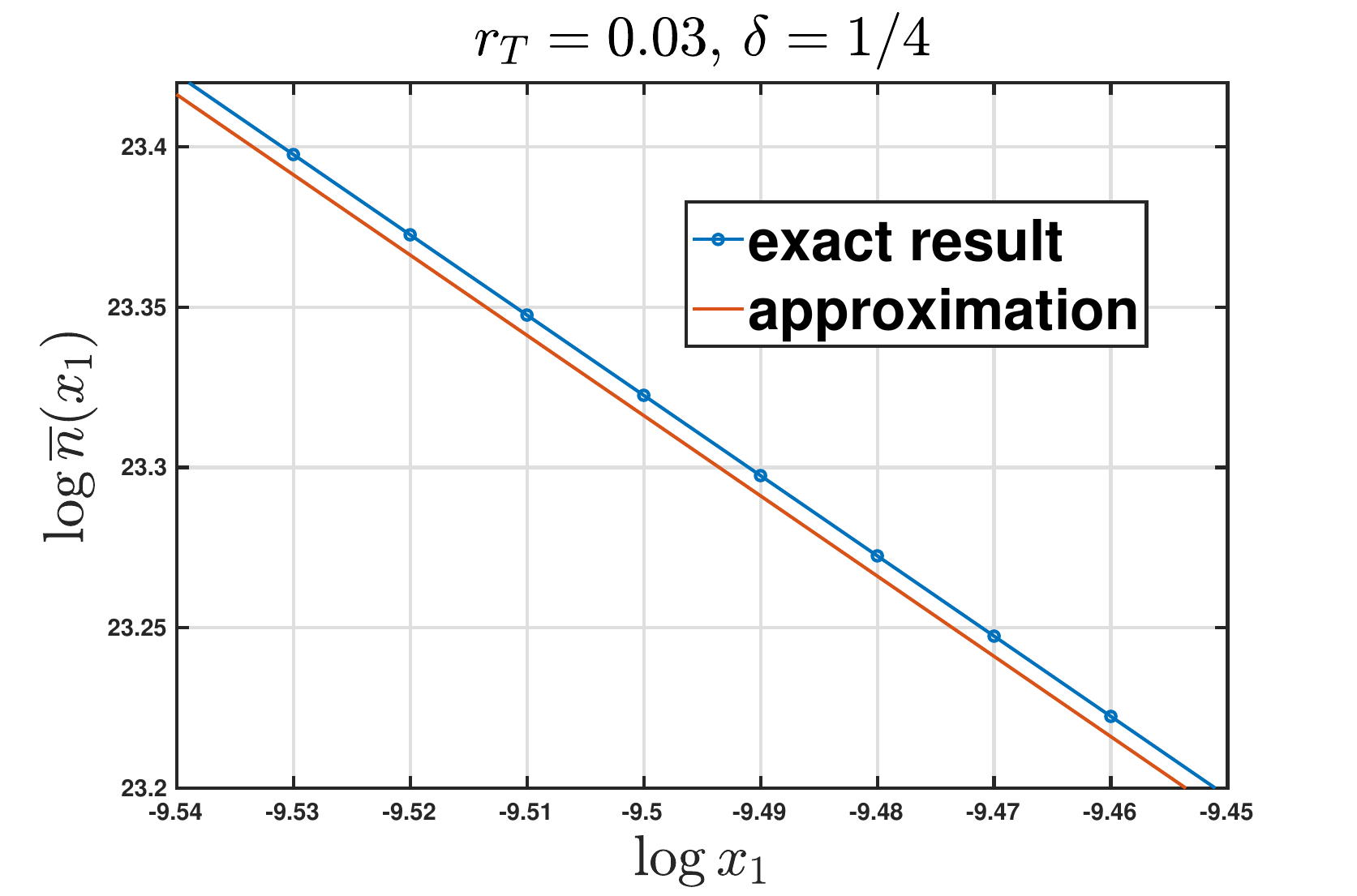}
\caption[a]{We compare the exact results based on Eq. (\ref{MULT3}) with the 
approximations leading to Eq. (\ref{MULT6}). Common logarithms are employed on both axes. The highly restricted 
range of variation of $x_{1}$ clarifies better than any other consideration the accuracy of the analytic approximation discussed in section \ref{sec5}. }
\label{FIGURE1}      
\end{figure}
In the limit $k \, y \gg 1$ and $x_{1}$ the averaged 
multiplicities are estimated for the different values of the parameters  with the same approximation schemes analyzed in section 
\ref{sec4}; from Eq. (\ref{MULT3}) we consider, in particular, the limit $k\, y \gg 1$ and deduce the momentum (or frequency) dependence 
of the averaged multiplicity 
\begin{equation}
\overline{n}(x_{1}) = \lim_{x_{1} <1,\,\,\,k\,y \gg 1} \, \overline{n}(x_{1}, k\, y) = 3\, {\mathcal B}( \delta, r_{T}) \, x_{1}^{n_{T}(\delta,r_{T}) -4},
\label{MULT6}
\end{equation}
where ${\mathcal B}(\delta,r_{T})$ has been already introduced in Eq. (\ref{SPEN2}) while $n_{T}(\delta, r_{T})$ follows from Eq. (\ref{SPIN1}). The analytic results of Eq. (\ref{MULT6}) are illustrated by the full line in both plots 
of Fig. \ref{FIGURE1}.  For a selection of parameters the analytic approximation is compared with the exact result of Eq. (\ref{MULT3}) where $k y_{0} = k y(\tau_{0}) = 10^{18}$. It can be checked that different typical values of $k \, y_{0}$ lead exactly to the same result as long as $k\, y_{0} \gg 1$. The results of Fig. \ref{FIGURE1} clarify the interplay between the analytic approximations and the scaling properties of the average multiplicity.
The estimates based on Eq. (\ref{MULT6}) differ from the exact results by factors that are ${\mathcal O}(10^{-2})$, as it is apparent from the scale of variation of $\overline{n}(x_{1})$ in both plots of Fig. \ref{FIGURE1}. In the limit $\delta \to 1$ the result of Eq. (\ref{MULT6}) reproduces the averaged multiplicity of the concordance scenario where an inflationary stage 
is replaced by an evolution dominated by radiation; in this case $n_{T} \to - r_{T}/8$ and 
$\overline{n}(x_{1}) \propto x_{1}^{-4 - r_{T}/8}$. In this last case the spectral energy density 
is therefore (approximately) scale-invariant and the average multiplicity 
is strongly non-thermal. 

Since the dimensionless variable $x_{1}$ is related to 
the ratio between $\nu$ and $\nu_{max}$, as the frequency scale increases there will necessarily be a region where  $\overline{n}(\nu, \tau_{0}) \to {\mathcal O}(1)$ and this is, from the operational viewpoint, the maximal frequency range of the spectrum. 
There are in fact two complementary ways in which $\nu_{max}$ can be determined \cite{MAX1}. 
As established  from Eqs. (\ref{FREQ2})--(\ref{FREQ3})) $\nu_{max}$
follows indirectly from the expansion rate after inflation and from the minimal value of the Hubble radius 
at the end of the inflationary stage. Conversely, in a quantum mechanical perspective, the maximal frequency of the spectrum corresponds to the production of few pairs of gravitons with opposite comoving three-momenta before the averaged 
multiplicity is exponentially suppressed
\begin{equation}
\lim_{x_{1} \to 1} \overline{n}(x_{1}) = {\mathcal O}(1), \qquad \nu \to \nu_{max}.
\label{MULT6a}
\end{equation}
Indeed, while the averaged multiplicity for $x_{1} < 1$ (i.e. $\nu <\nu_{max}$) corresponds to the mean number of produced pairs  for $x_{1}>1$ the averaged multiplicity is suppressed  exponentially or, more precisely \cite{AA01,AA02,BIRREL},
\begin{equation}
\frac{\bigl|\overline{n}(x_{1}) \bigr|^2 }{1 + \bigl|\overline{n}(x_{1}) \bigr|^2 } = e^{- \gamma\,x_{1}}, \qquad\qquad x_{1} >1.
\label{MULT6b}
\end{equation}
The degree of suppression of Eq. (\ref{MULT6b}) depends on $\gamma$, i.e. a numerical factor ${\mathcal O}(1)$ controlled by the smoothness of the transition between the 
inflationary and the post-inflationary phase; the value of $\gamma$ can be numerically estimated if the evolution of the mode functions is carefully integrated frequency by frequency \cite{mg0}. By looking simultaneously at Eqs. (\ref{MULT6}) and (\ref{MULT6b}) the averaged 
multiplicity of pairs produced from the vacuum can be written in the following suggestive form:
\begin{equation}
\overline{n}(x_{1}) = \frac{3 \gamma \, {\mathcal B}(\delta, r_{T})  \,\, x_{1}^{n_{T} - 3}}{e^{\gamma\,x_{1}} -1},\qquad\qquad x = (\nu/\nu_{max}).
\label{MULT6c}
\end{equation}
Equation (\ref{MULT6c})
interpolates between the power-law behaviour of Eq. (\ref{MULT6}) and the exponential suppression of Eq. (\ref{MULT6b}). This means that  the spectrum of relic gravitons is in fact 
 a distorted thermal spectrum as argued long ago \cite{PARKTH}; this conclusion is incidentally consistent with the multiplicity distribution of the relic gravitons which is of Bose-Einstein type \cite{MAX1}. 
\begin{figure}[!ht]
\centering
\includegraphics[height=6.5cm]{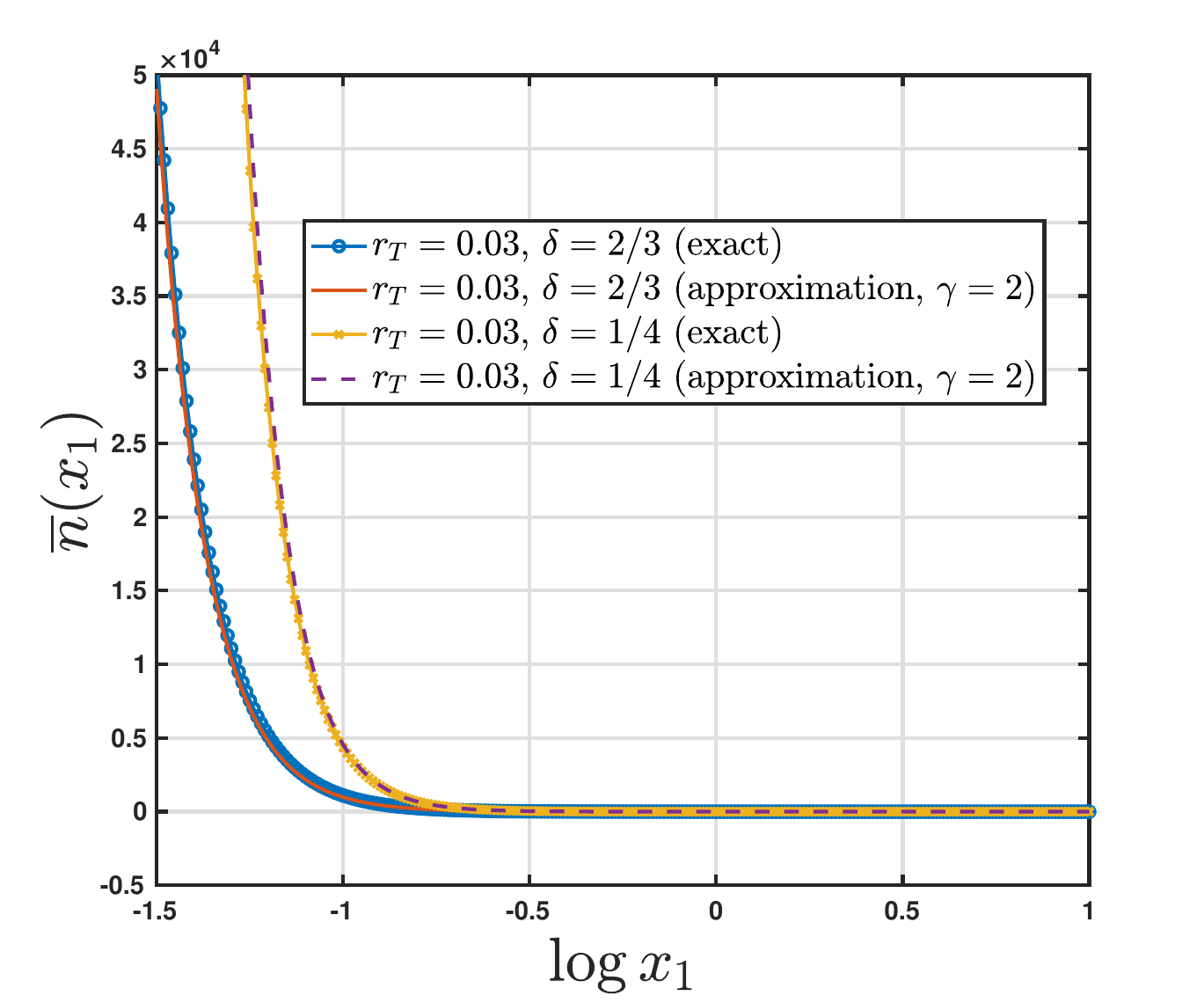}
\includegraphics[height=6.5cm]{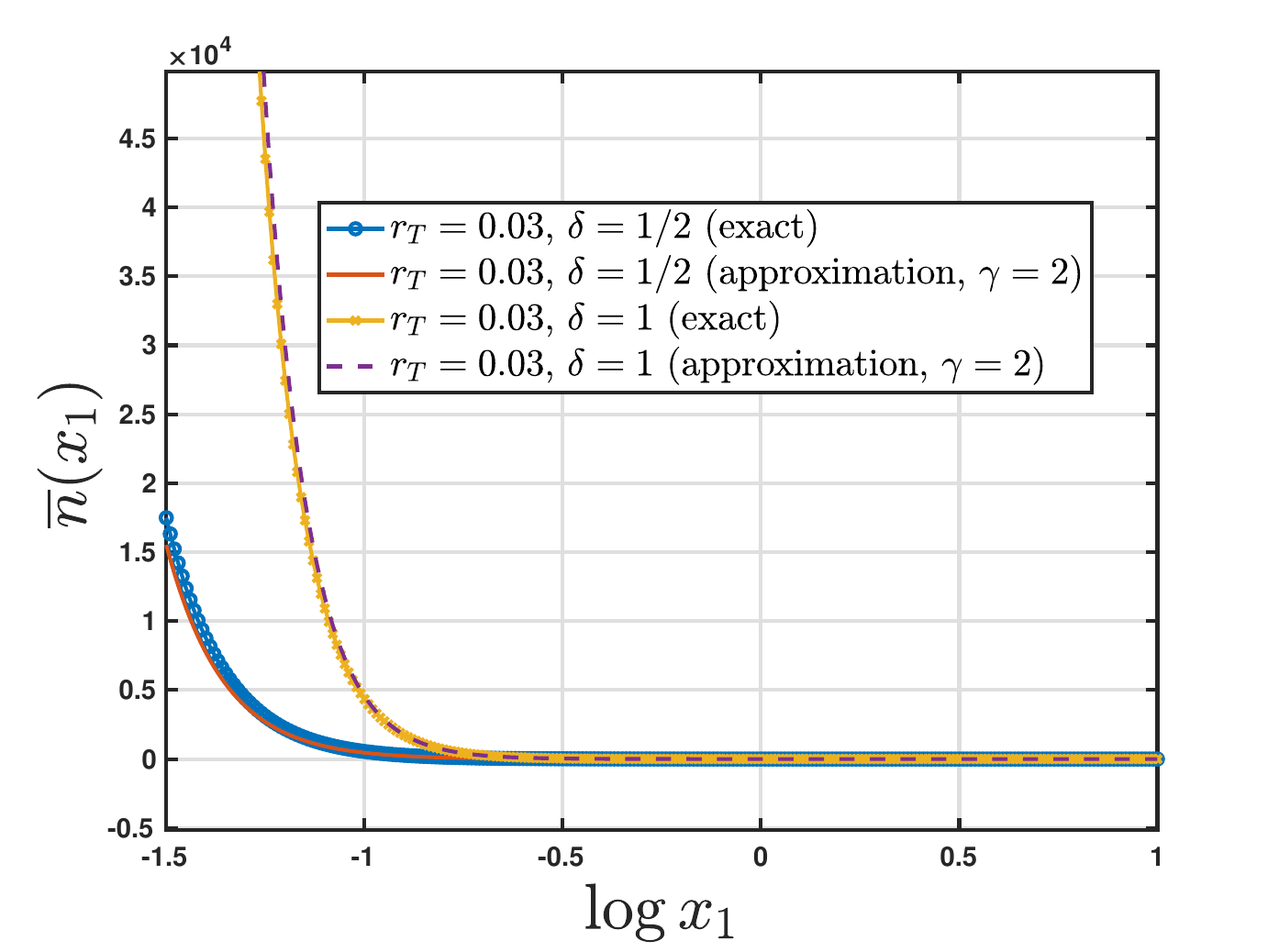}
\caption[a]{In both plots we compare the exact results based on Eq. (\ref{MULT3}) with the 
approximations discussed in Eq. (\ref{MULT6c}). Unlike the case of Fig. \ref{FIGURE1}, to ease the comparison and to clarify the exponential suppression, the average multiplicity has been illustrated on a linear scale as a function of the common logarithm of $x_{1}$.  We can also appreciate from both plots that when $\gamma \to 2$ the exact result is well represented by the analytic approximation.}
\label{FIGURE2}      
\end{figure}
The exact result of Eq. (\ref{MULT3}) is illustrated in Fig. \ref{FIGURE2} and compared with the interpolation of Eqs. (\ref{MULT6b})--(\ref{MULT6c}). We selected the same value of $k \, y_{0}$ employed in Fig. \ref{FIGURE1}; as previously remarked, as long as  limit $k\,y_{0}\gg 1$ the specific value is immaterial. In Fig. \ref{FIGURE2} we considered 
the range around the maximal frequency corresponding to $x_{1} = {\mathcal O}(1)$ and from the  
comparison illustrated in Fig. \ref{FIGURE2} we can infer that Eq. (\ref{MULT6c}) 
reproduces the exact results for $\gamma = {\mathcal O}(2)$. It is known 
that the specific amount of exponential suppression depends on the nature of the transition 
regime that has been numerically discussed in Refs. \cite{mg0}. It is 
however interesting that the minimal framework leading to Eq. (\ref{MULT3}) (based 
on the continuity of the transition matrix and of the background) generally suggests $\gamma = {\mathcal O}(2)$.
As we are going to see in a moment, the exponential suppression for $\nu>\nu_{max}$  
also guarantees the convergence of the integrated spectral energy density over the whole 
frequency domain \cite{mg0} (see also \cite{MAX1}).

\subsection{Scaling of the spectral energy density} 
The averaged multiplicity $\overline{n}(x_{1}, k\, y)$ can be related to  the rescaled spectral energy density in critical units denoted hereunder by $\omega_{gw}(x_{1}, k\, y)$: 
\begin{equation}
\omega_{gw}(x_{1}, k\, y) = x_{1}^4 \,\,\overline{n}(x_{1}, k\, y).
\label{MULT4}
\end{equation}
This analysis is useful for numerical purposes  but  $\omega_{gw}(x_{1}, k\, y)$ is simply related to the value of 
$\Omega_{gw}(x_{1}, k\, y)$; more specifically the relation between the two parametrizations does not depend 
on the frequency and it is given by:
\begin{equation}
\Omega_{gw}(x_{1}, k\, y) = \frac{8 }{3 \pi} \biggl(\frac{H_{1}}{M_{P}}\biggr)^2 \biggl( \frac{H_{1}^2 \, a_{1}^4}{H^2 \, a^4} \biggr) \, \omega_{gw}(x_{1}, k\, y).
\label{MULT5}
\end{equation}
It the follows from the result of Eq. (\ref{MULT6}) that $\omega_{gw}(x_{1}, k\, y)$ should scale as 
\begin{equation}
 \lim_{x_{1} <1,\,\,\,k\,y \gg 1} \, \frac{\omega_{gw}(x_{1}, k\, y)}{x_{1}^{n_{T}}} = 3\, {\mathcal B}( \delta, r_{T}),
 \label{MULT7}
 \end{equation}
so that the ratio $\omega_{gw}(x_{1}, k\, y)/x_{1}^{n_{T}}$ does not depend on $x_{1}$ 
but only on the specific values of $r_{T}$ and $\delta$ encoded in the expression of 
${\mathcal B}(\delta, r_{T})$.
\begin{figure}[!ht]
\centering
\includegraphics[height=6.6cm]{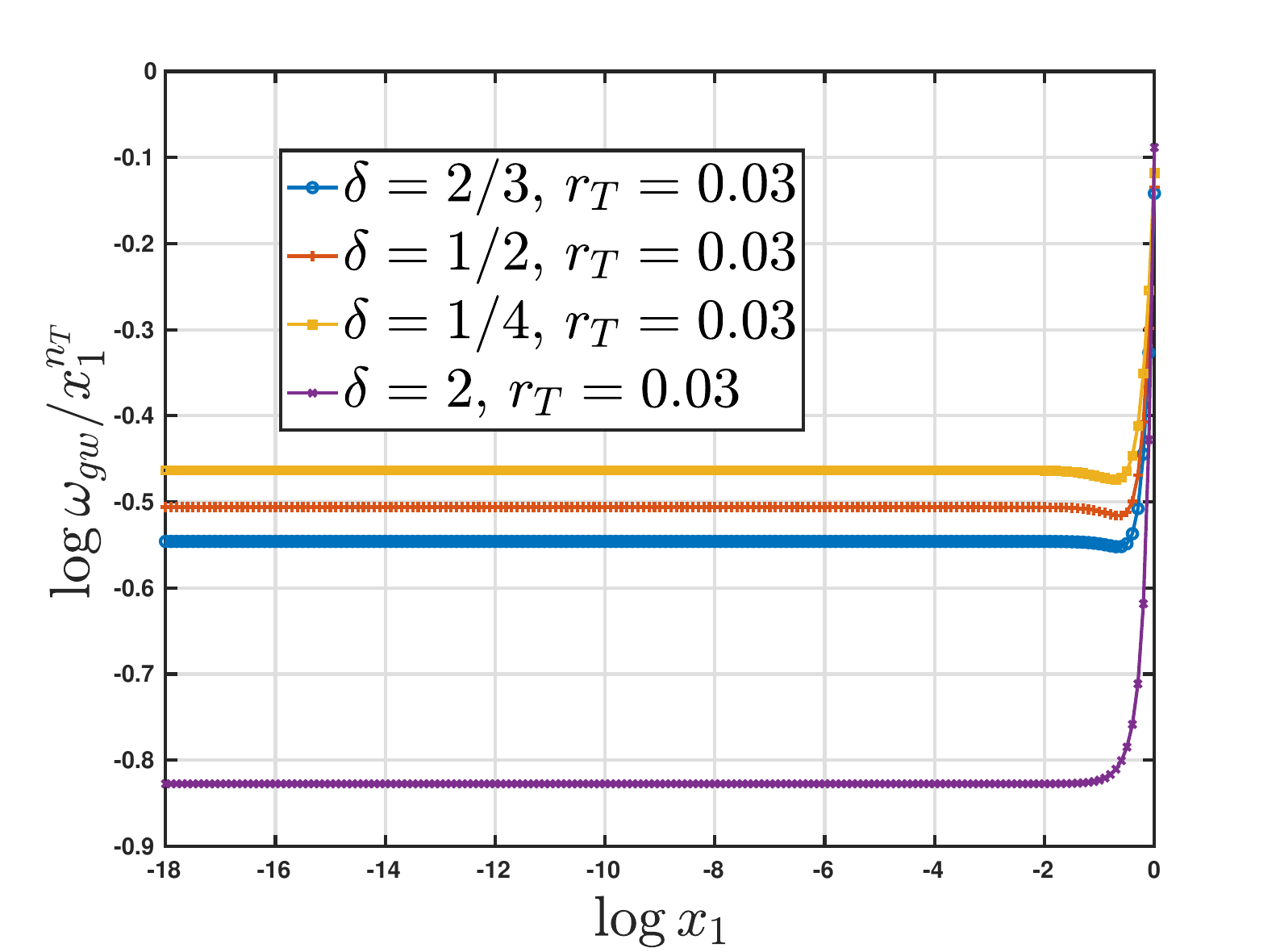}
\includegraphics[height=6.6cm]{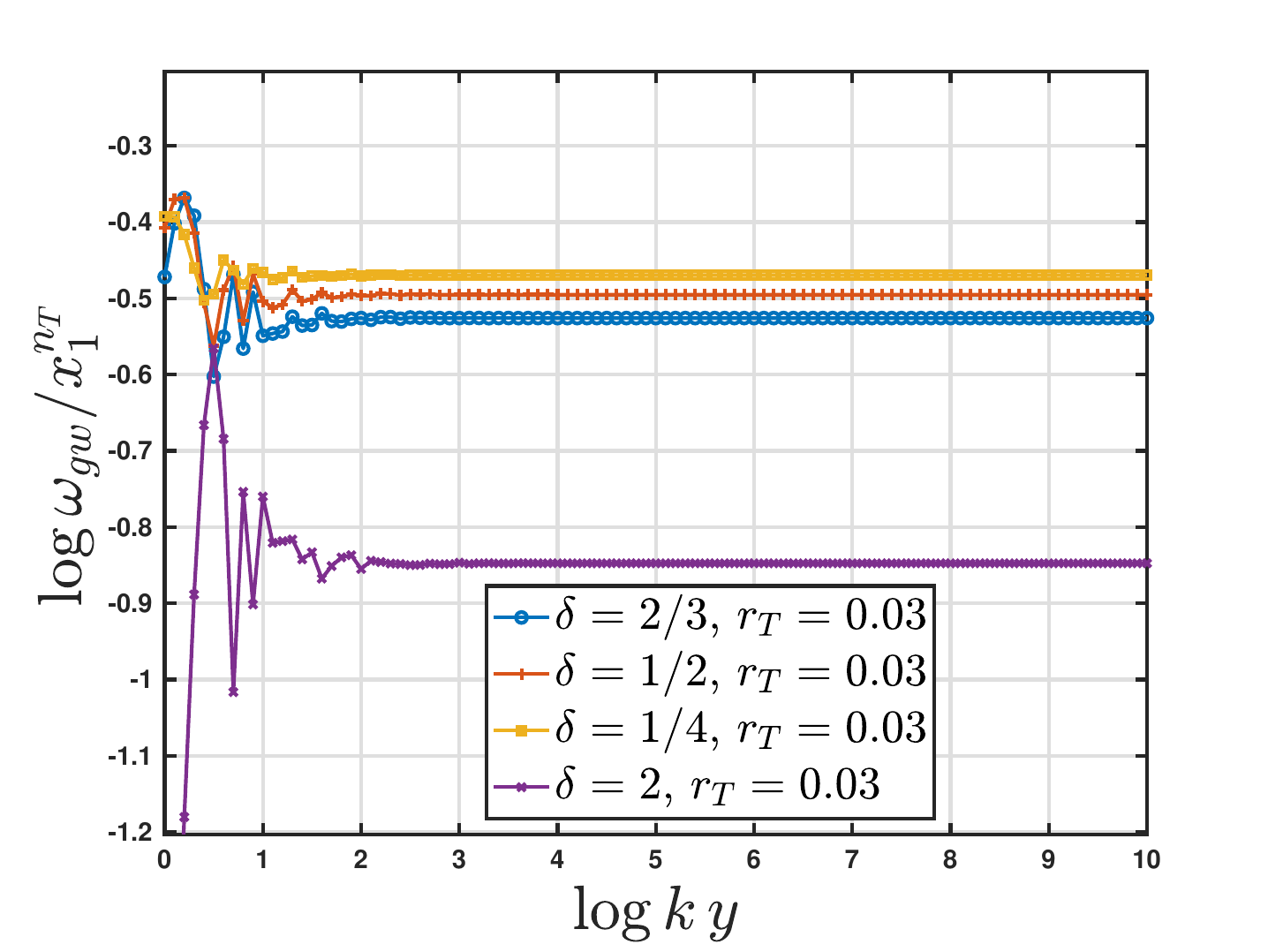}
\caption[a]{In the left plot the common logarithm of $\omega_{gw}(x_{1}, k\, y)/x_{1}^{n_{T}}$ is illustrated 
as a function of the common logarithm of $x_{1}$ when $r_{T} = 0.03$ and for different values of $\delta$. Although the value of $k\,y$ is immaterial (as long as $k \,y >1$) the left plot has been obtained for $k y = 10^{15}$. In the right plot the value of $x_{1}$ has been fixed to $10^{-8}$ and the common logarithm of $\omega_{gw}(x_{1}, k\, y)/x_{1}^{n_{T}}$ has been illustrated 
as a function of the common logarithm of $k\, y$.}
\label{FIGURE3}      
\end{figure}
In Fig. \ref{FIGURE3} we illustrate the common logarithm of the 
ratio $\omega_{gw}(x_{1}, k\, y)/x_{1}^{n_{T}}$ as a function of the common logarithm of $x_{1}$. 
As expected from Eq. (\ref{MULT7}) the results of Fig. \ref{FIGURE2} are scale-invariant
and the explicit values of the plateau are correctly estimated by the analytic approximation 
with the same accuracy illustrated by Fig. \ref{FIGURE1}. This means 
that the analytic expressions of the spectral energy density in critical units 
can be used to infer the chirp amplitudes and the potential signals in different 
frequency ranges.

\begin{figure}[!ht]
\centering
\includegraphics[height=6.2cm]{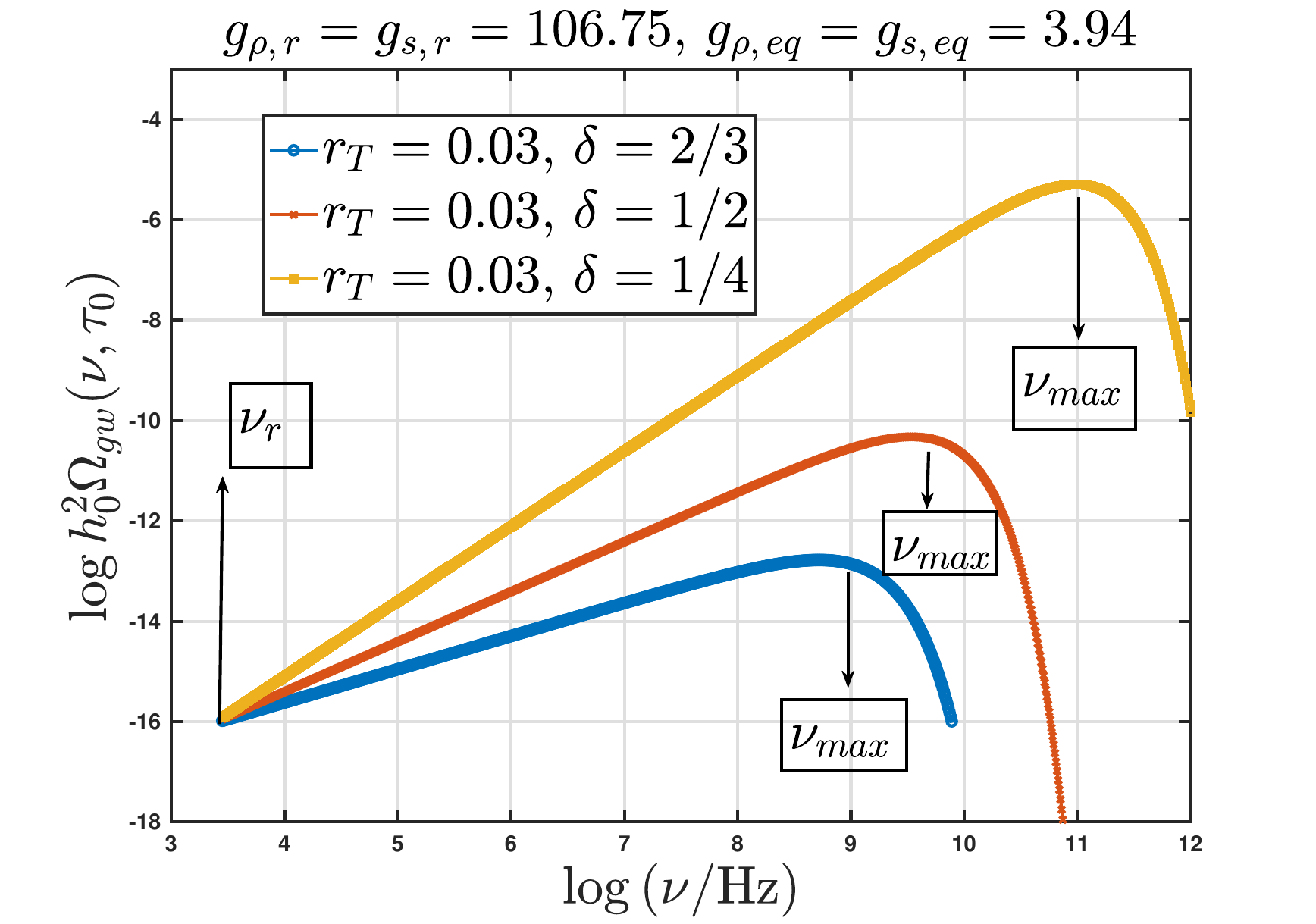}
\includegraphics[height=6.2cm]{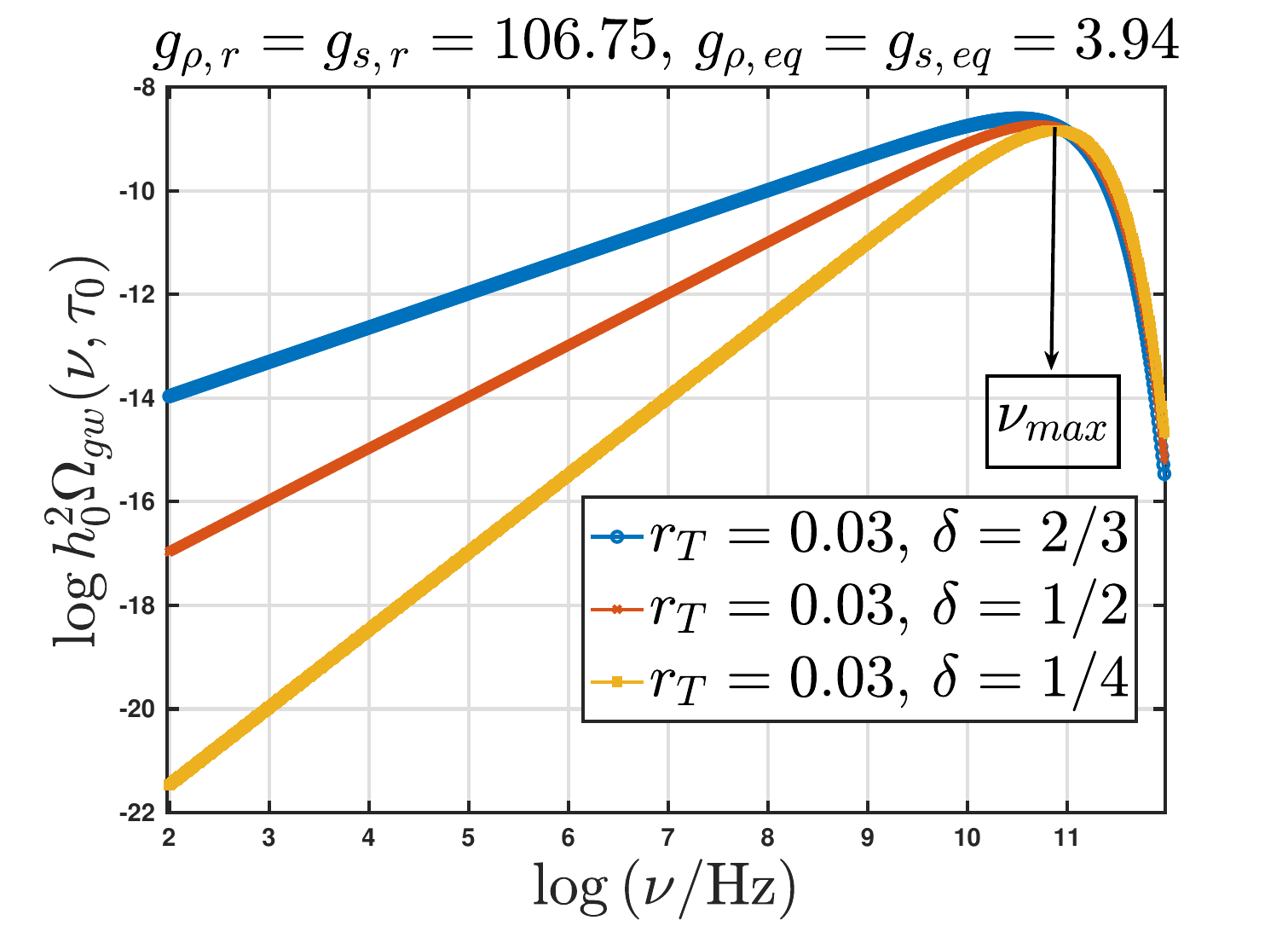}
\caption[a]{In the left plot the spectral energy density in critical units computed from Eqs. (\ref{SIXEEc}) and (\ref{SIXEEca}) is approximately normalized at $\nu_{r}$, i.e. the lowest frequency compatible with 
the analytic treatment (see also Eq. (\ref{nur1})); different selections of the parameters are illustrated. In the right plot we instead normalize the same curves in the highest frequency domain compatible with the bounds of Eqs. (\ref{FF2}) and (\ref{HH3}).}
\label{FIGURE4}      
\end{figure}

The present analysis suggests that the high frequency and ultra-high frequency signals can be estimated 
according to a twofold strategy. In the first approach the potential signal is normalized well below the maximal frequency 
whereas, in the second possibility, the spectral energy density is directly fixed at the maximal frequency; this twofold strategy is illustrated in Fig. \ref{FIGURE4}. In the realistic situation the analytic approximation 
discussed so far only makes sense for sufficiently large frequencies 
\begin{equation}
\nu > \nu_{r} = \frac{(2 \Omega_{R0})^{1/4}}{2 \pi} \, \sqrt{H_{0}\,  H_{r}} \biggl(\frac{g_{\rho,\, r}}{g_{\rho,\,eq}}\biggr)^{1/4} \biggl(\frac{g_{s,\, eq}}{g_{s,r}}\biggr)^{1/3},
\label{nur1}
\end{equation}
where $H_{r}$ now denotes the value of the expansion rate at the onset of the radiation-dominated stage. From Eq. (\ref{nur1}) the explicit value of $\nu_{r}$ is related to $\overline{\nu}_{max}$ since, by definition, $\nu_{r}/\overline{\nu}_{max} = \sqrt{H_{r}/H_{1}}$. One possibility is therefore to normalize the spectral energy 
density in critical units for $\nu \to \nu_{r}$, as illustrated in the left plot of Fig. \ref{FIGURE4}.
A complementary strategy is to match $h_{0}^2 \Omega_{gw}(\nu, \tau_{0})$ with its maximally allowed value at the frequency of the spike, as 
suggested in the right plot of Fig. \ref{FIGURE4}. In this case, however, the maximum 
of the spectral energy density should be determined in absolute terms \cite{MAX1}. In Fig. \ref{FIGURE4} we have set this 
absolute value to be ${\mathcal O}(0.1)$ THz.  

It is appropriate to mention, at this point, that the timeline of the expansion rate can be more 
complicated than the one considered so far. In this case all the considerations discussed here 
in the case of a single post-inflationary stage can be extended exactly with the same techniques 
by following the considerations developed, at length, in Ref. \cite{MGrev}. 
Depending 
on the number of the intermediate stages also the typical frequencies of the spectrum increase. Rather than starting from the general considerations it is better to consider a specific example. For instance in the case of two successive intermediate phases 
with expansion rates $\delta_{1}$ and $\delta_{2}$ we have that $\nu_{max}$ becomes
\begin{equation}
\nu_{max} = \nu_{1} = \prod_{i\,=\,1}^{2} \, \xi_{i}^{\frac{\delta_{i} -1}{2 (\delta_{i}+1})}\,\, \overline{\nu}_{max} = \xi_{1}^{\frac{\delta_{1} -1}{2 (\delta_{1}+1})}\, \xi_{2}^{\frac{\delta_{2} -1}{2 (\delta_{2}+1})}\,\, \overline{\nu}_{max},
\label{INTT1}
\end{equation}
where $\xi_{1} = H_{2}/H_{1}$ and $\xi_{2} = H_{r}/H_{2}$.
The intermediate frequencies $\nu_{2}$ and $\nu_{r}$ are related to $\overline{\nu}_{max}$ and 
they are
\begin{eqnarray}
\nu_{2} &=& \sqrt{\xi_{1}} \,\, \xi_{2}^{(\delta_{2} -1)/[2 (\delta_{2}+1)]} \, \,\overline{\nu}_{max},
\nonumber\\
\nu_{r} &=&\nu_{3} = \sqrt{\xi_{1}}\, \sqrt{\xi_{2}}\, \overline{\nu}_{max} = \sqrt{\xi_{r}} \,\overline{\nu}_{max},
\label{INTT2}
\end{eqnarray}
where, by definition, $\xi_{r} = \, \xi_{1} \, \xi_{2} = H_{r}/H_{1}$. Both Eqs. (\ref{INTT1}) and (\ref{INTT2}) 
can be generalized to the case of $n$ intermediate stages of expansion \cite{MGrev}.

\subsection{The ultra-high frequency and the low frequency normalizations}
In Eq. (\ref{FF2}) it has been suggested that the upper bound on the frequency 
should be indeed of the order of the THz and this is what matters to set the absolute 
normalization in the ultra-high frequency regime. From a classical viewpoint the maximal frequency 
coincides with the minimal amplified wavelength of the spectrum and it ultimately 
depends on the post-inflationary evolution. 
This is, in a nutshell, the approach 
leading to Eqs. (\ref{FREQ2})--(\ref{FREQ3}): $\nu_{max}$ gets modified depending 
upon the post-inflationary evolution.  Equations (\ref{FREQ2})--(\ref{FREQ3}) assume a
single post-inflationary stage for curvature scales $H> H_{r}$ and this is why  
$\nu_{max}$ depends on $\delta$, $H_{1}$ and $H_{r}$. A consequence of this observation is, for instance, 
that for the same value of the ratio $H_{r}/H_{1} <1$ the frequency $\nu_{max}$ is either larger or smaller than 
${\mathcal O}(100)$ MHz depending if $\delta < 1$ or $\delta >1$. 
\begin{figure}[!ht]
\centering
\includegraphics[height=6.6cm]{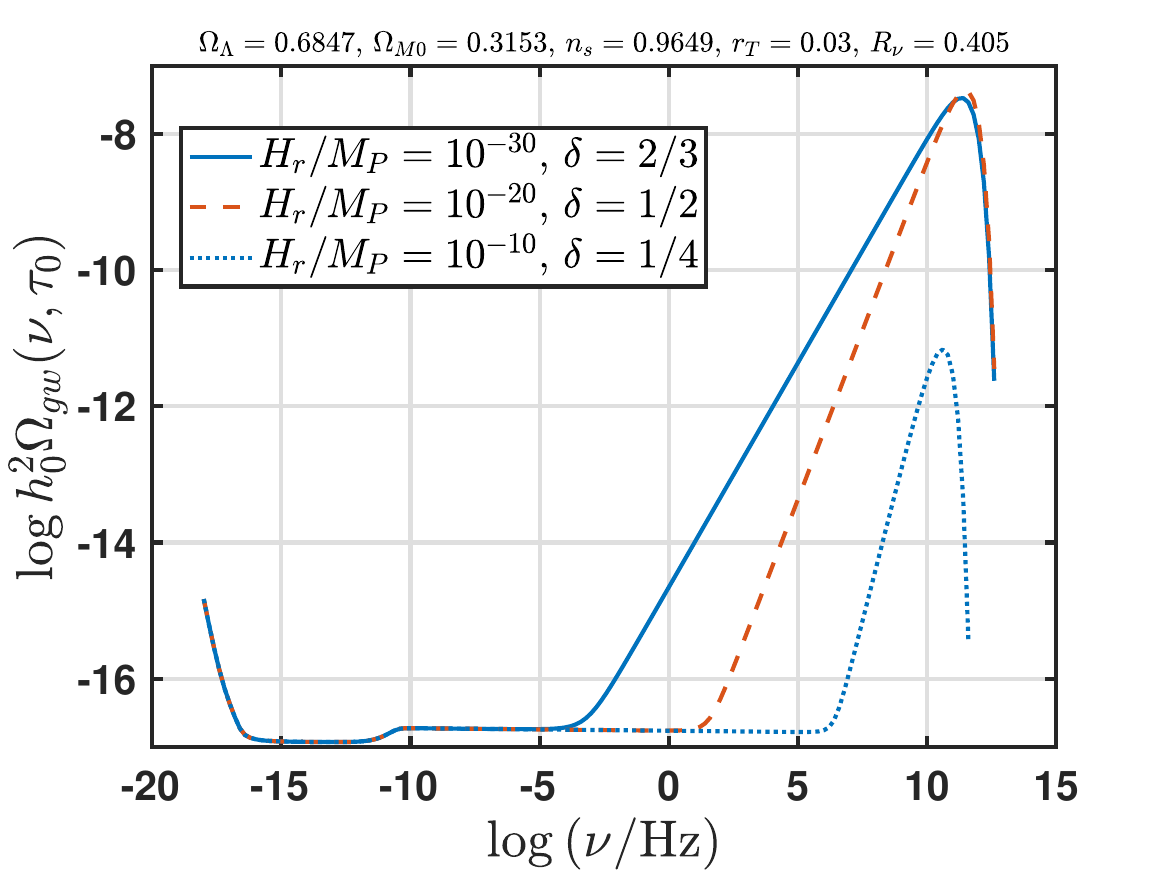}
\includegraphics[height=6.6cm]{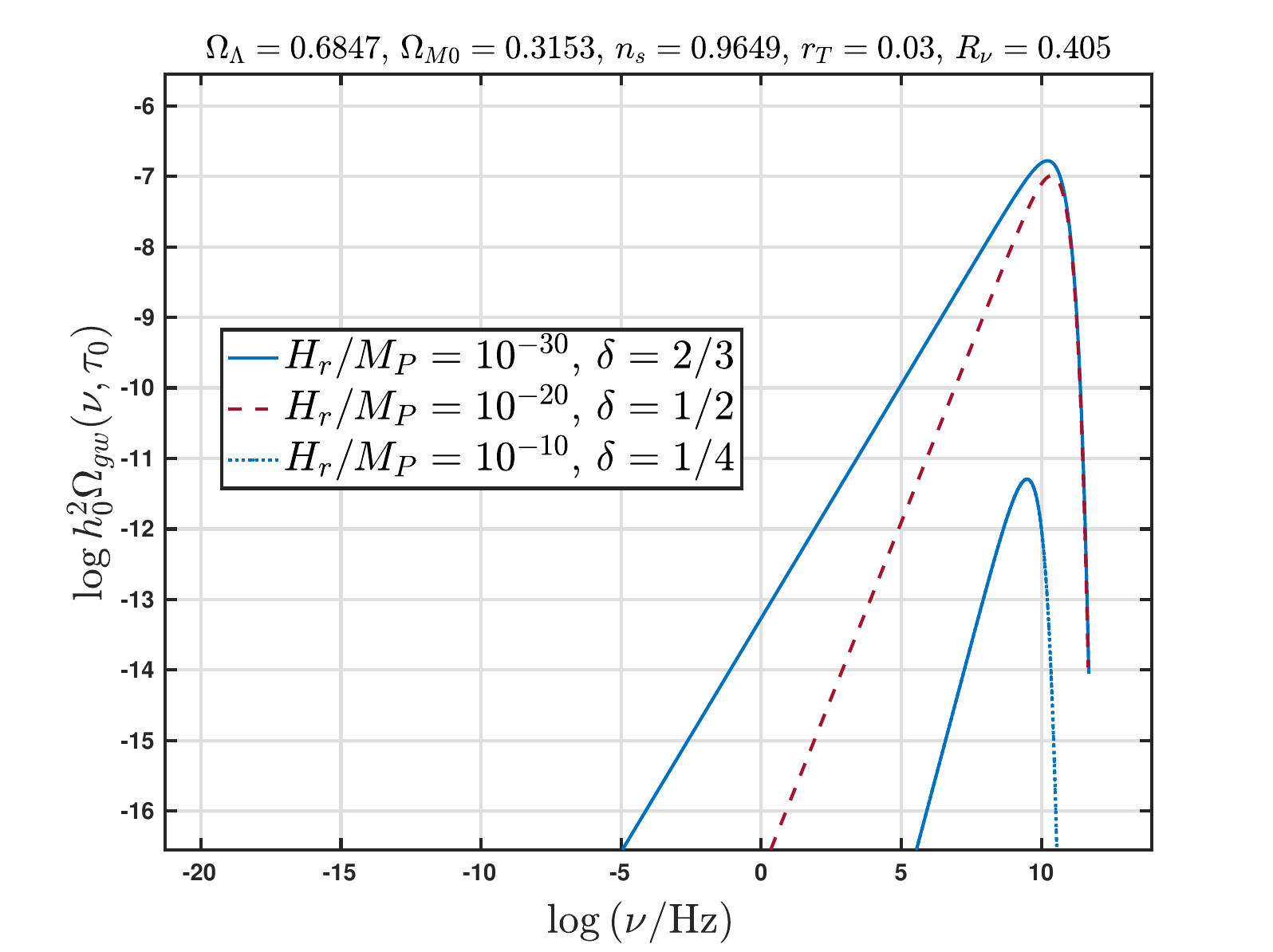}
\caption[a]{In the left plot the common logarithm of $h_{0}^2 \Omega_{gw}(\nu,\tau_{0})$ is illustrated as a function of the common logarithm of the frequency. For both plots the post-inflationary expansion rate between $H_{1}$ and $H_{r}$ is slower than radiation;
the parameters involving the post-inflationary stage 
are the same in both plots. In the left plot the suppression below the nHz region comes predominantly from the 
free-streaming of the neutrinos. In the right plot we illustrate instead the analytic approximation (see  Eqs. (\ref{SIXEEc}) and (\ref{SIXEEca})) normalized in the high frequency domain. }
\label{FIGURE5}      
\end{figure}
Along a purely quantum mechanical perspective it is instead possible to derive an absolute upper bound on $\nu_{max}$ and while this analysis has been already 
presented in the recent past \cite{MAX1}, it is now useful to deduce this bound in a slightly 
different manner. The first step is to express the spectral energy density in critical units in terms 
of the averaged multiplicity of the produced gravitons with opposite three-momenta: 
\begin{equation}
\Omega_{gw}(\nu, \tau) = \frac{128\, \pi^3}{3} \,\, \frac{\nu^{4}}{H^2 \, M_{P}^2\, a^4}\,\, \overline{n}(\nu, \tau).
\label{ABS1}
\end{equation}
 Equation (\ref{ABS1}) suggests 
that the maximal frequency corresponds to the production a single pair relic gravitons: for frequencies 
smaller than $\nu_{max}$ the averaged multiplicity scales like a power-law while $\overline{n}(\nu, \tau_{0})$ 
is exponentially suppressed for $\nu > \nu_{max}$. When only one pair is produced we have, from Eq. (\ref{ABS1}),
\begin{equation}
  \Omega_{gw}(\nu, \tau_{0}) = \frac{128\, \pi^3}{3} \,\, \frac{\nu_{max}^{4}}{H_{0}^2 \, M_{P}^2}. 
  \label{ABS1a}
\end{equation}
Equation (\ref{ABS1a}) sets already a generous limit on $\nu_{max}$ 
since $H_{0} = 3.24 \,h_{0}\, \mathrm{aHz}$ (we remind that $1\, \mathrm{aHz} = 10^{-18}\, \mathrm{Hz}$) and $M_{P}= 1.855\times10^{31}\, \mathrm{THz}$: from the condtion  $\Omega_{gw}(\nu, \tau_{0})< 1$  the maximal frequency must be already of the order of the THz but there is in fact a more constraining condition since the wavelengths reentering the Hubble radius between the end of inflation and big bang nucleosynthesis must comply with the bound
\begin{equation}
h_{0}^2 \, \int_{\nu_{bbn}}^{\nu_{max}} \,\Omega_{gw}(\nu,\tau_{0}) \,\,d\ln{\nu} < 5.61\times 10^{-6} \biggl(\frac{h_{0}^2 \,\Omega_{\gamma0}}{2.47 \times 10^{-5}}\biggr) \, \Delta N_{\nu},
\label{ABS2}
\end{equation}
where $\Omega_{\gamma 0}$ is the (present) critical fraction of CMB photons and the integral at the left hand side 
of Eq. (\ref{ABS2}) must be evaluated, as indicated, between the frequency of big bang nucleosynthesis (i.e. $\nu_{bbn}$) and 
$\nu_{max}$.  

Equation (\ref{ABS2}) sets an indirect constraint on the potential presence of extra-relativistic species at the time of nucleosynthesis \cite{DD2,DD3,DD4}. Since Eq. (\ref{ABS2}) is also relevant in the context of neutrino physics the limit is often expressed in terms of $\Delta N_{\nu}$ (i.e. the contribution of supplementary neutrino species). The constraints on $\Delta N_{\nu}$ range from $\Delta N_{\nu} \leq 0.2$ to $\Delta N_{\nu} \leq 1$ so that the integrated spectral density in Eq. (\ref{ABS2}) must range, at most, between  $10^{-6}$ and $10^{-5}$. Given the expansion rate at the big bang nucleosynthesis time (when the temperature of the plasma was 
approximately ${\mathcal O}(1)$ MeV),  $\nu_{bbn}$ can be expressed as:
\begin{eqnarray}
\nu_{bbn} &=& 8.17 \times 10^{-33} g_{\rho,\, bbn}^{1/4} \, T_{bbn} \biggl(\frac{h_{0}^2 \Omega_{R0}}{4.15\times 10^{-5}} \biggr)^{1/4}
\nonumber\\
&=& {\mathcal O}(2) \times 10^{-2} \biggl(\frac{g_{\rho,\, bbn}}{10.75}\biggr)^{1/4} \biggl(\frac{T_{bbn}}{\,\,\mathrm{MeV}}\biggr) 
\biggl(\frac{h_{0}^2 \Omega_{R0}}{4.15 \times 10^{-5}}\biggr)^{1/4}\,\,\mathrm{nHz}.
\label{ABS3}
\end{eqnarray}

In a complementary perspective the diffuse background of relic gravitons was produced 
after BBN. In this case we observe that $\Omega_{gw}(\nu, \tau_{0})$ 
must always be smaller than the the current fraction of relativistic species to avoid an observable impact on the CMB and matter power spectra \cite{DD5,DD6}. The two bounds are actually numerically compatible so that the final limit on $\nu_{max}$ is given by:
\begin{equation}
\nu_{max} \leq 0.165 \,\, \Omega_{R0}^{1/4}\,\, \sqrt{H_{0}\, M_{P}} < \,\mathrm{THz},
\label{HH3}
\end{equation}
where the inequality follows in the limit $\overline{n}(\nu_{max}, \tau_{0}) \to {\mathcal O}(1)$, i.e. in the case where a single pair of gravitons is produced. The bound of Eq. (\ref{HH3}) is numerically more restrictive than the bounds derived from Eqs. (\ref{FREQ2}) and (\ref{FREQ3}) on the basis of classical considerations. We also stress that in Fig. \ref{FIGURE4} the upper curve of the 
left plot is not obviously compatible with the the bound of Eq. (\ref{ABS2}). This means 
that if we want to avoid the analysis of the low frequency domain of the spectrum, the 
analytic results for the spike should be mandatorily normalized by using the approach 
described in this subsection. The conclusions obtained from this strategy should not be interpreted 
as predictions but rather as upper bounds. 

We are now going to consider more closely the interplay between the analytic results illustrated  
in the first part of this section and the numerical evaluation appropriately
normalized at low frequencies \cite{MAX2,MAX3}. In this 
respect the first point to be stressed is that 
between the nHz domain and the audio band (with  $\mathrm{Hz} < \nu_{audio} < 10\,  \mathrm{kHz}$)  the spectral energy density of inflationary origin is, at most, ${\mathcal O}(10^{-17})$ in critical units and  the deviations from scale-invariance in the direction of blue spectral indices are excluded at least in the conventional situation where the corrections to $h_{0}^2\,\Omega_{gw}(\nu,\tau_{0})$ always lead to decreasing spectral slope \cite{TT1,TT2,TT3}. The second relevant aspect 
is that, in the case of the concordance paradigm the spectral energy density is further reduced by various sources of damping and, most notably, by the free-streaming of neutrinos \cite{STRNU0,STRNU1,STRNU2,STRNU3,STRNU4} operating below a fraction of the nHz. The same phenomenon also affects the spectral energy density when the corresponding slopes are increasing \cite{mg0}; in both situations, however, the suppression due to the neutrinos operates for $\nu < \nu_{bbn}$ and when the expansion rate is already dominated by radiation.
In Fig. \ref{FIGURE5} we illustrate a comparison between the analytic approximation
normalized at the corresponding maximal frequency (see Eqs. (\ref{FREQ2})--(\ref{FREQ3}))
and the full spectrum normalized at low frequency. The high frequency parameters of both plots are the same and we can see that, overall, the two results coincide within one order of magnitude both 
in the frequency domain and in the amplitude. In the aHz region the spectral energy density decreases as $\nu^{-2}$ while we can appreciate the suppression due to the neutrino free streaming close to $\nu_{bbn}$ \cite{STRNU0,STRNU1,STRNU2,STRNU3,STRNU4}. Other sources of suppression taken into account in Fig. \ref{FIGURE5} and in the remaining plots include the late-time dominance of dark energy and the evolution of relativistic species. The spectra of Fig. \ref{FIGURE5} have been deduced by using 
for the fiducial parameters the last Planck data release in the case of three massless neutrinos where $R_{\nu} = \rho_{\nu}/(\rho_{\gamma} + \rho_{\nu}) =0.405$, as indicated on top of each plots; this is the choice of the minimal $\Lambda$CDM scenario. If the radiation would dominate the whole post-inflationary evolution the quasi-flat plateau (decreasing because of the slow-roll corrections) would last up to frequencies ${\mathcal O}(300)$ MHz.

\subsection{Absolute bounds on the post-inflationary expansion history}
The analytic results Eqs. (\ref{SIXEEc}) and (\ref{SIXEEca}) leading to the right plot of Fig. \ref{FIGURE5} can also be employed to obtain more swiftly all the relevant 
bounds on the duration on the post-inflationary expansion rate.   
 \begin{figure}[!ht]
\centering
\includegraphics[height=7.4cm]{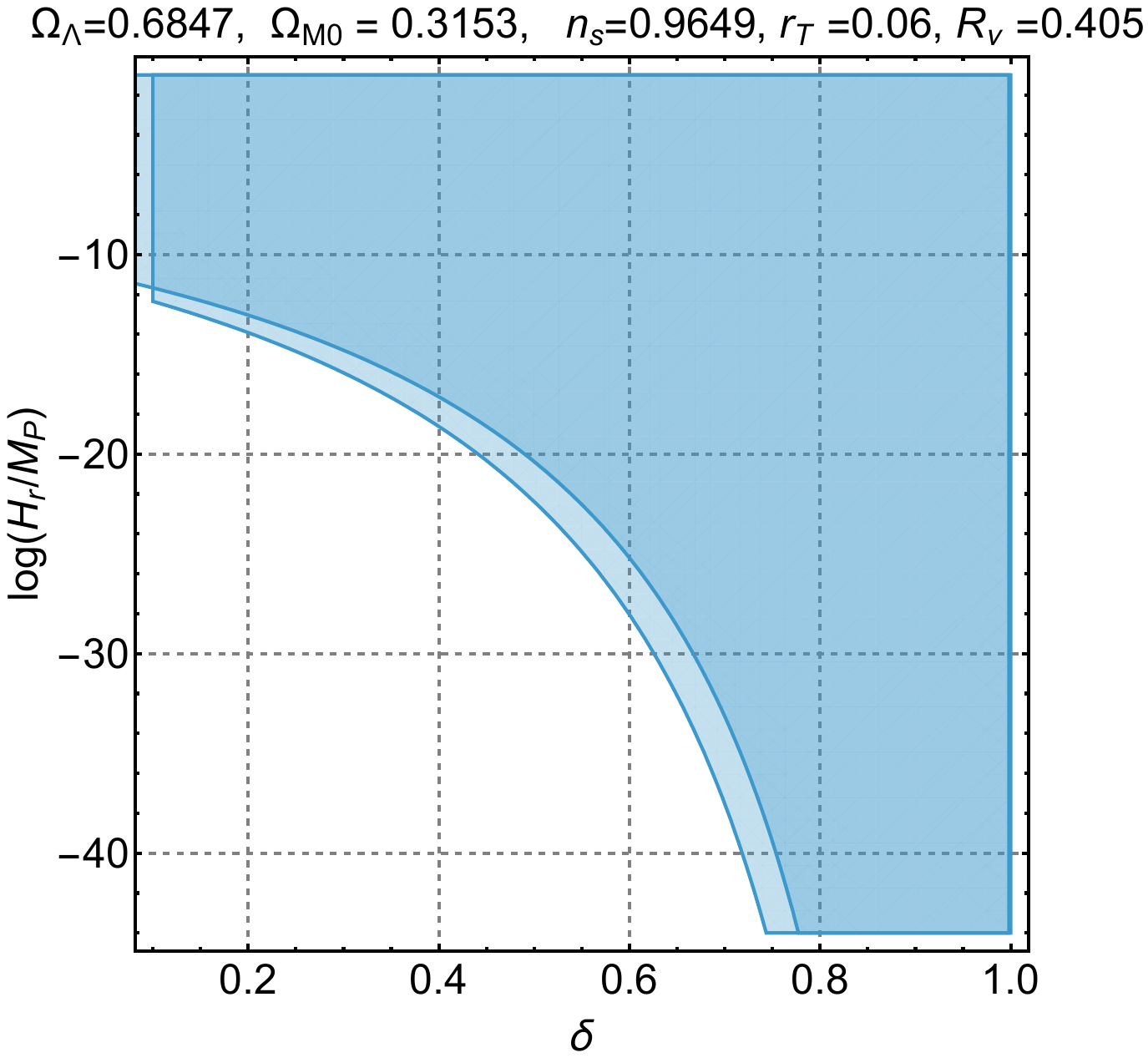}
\includegraphics[height=7.4cm]{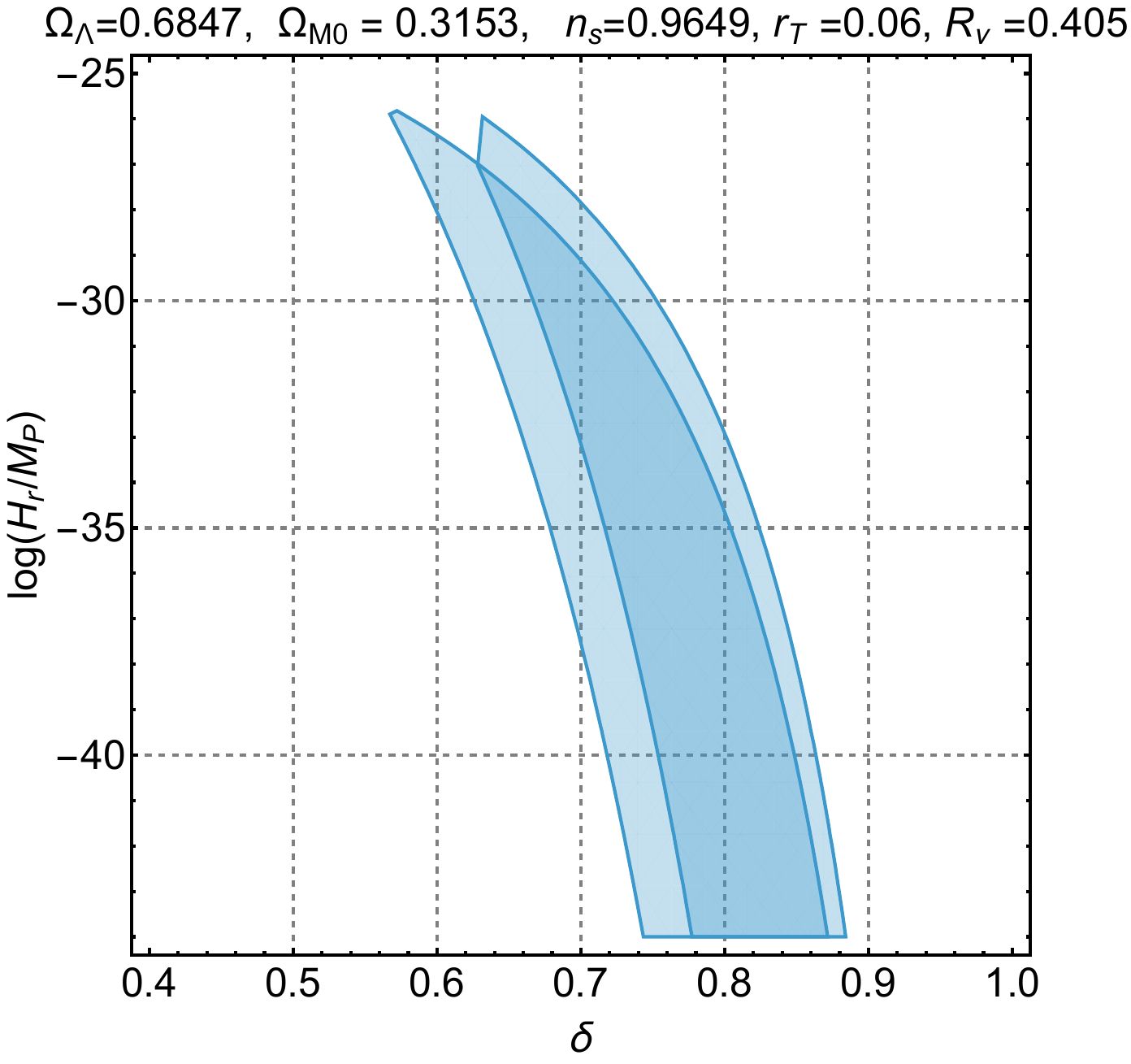}
\caption[a]{In the left plot the bound of Eq. (\ref{ABS2}) is illustrated in the plane 
$(H_{r}/M_{P}, \,\delta)$. The two shaded regions correspond to the approaches already 
scrutinized in Fig. \ref{FIGURE5}.  In the plot at the left the shaded areas denote the regions compatible with the BBN limit; in the 
right plot the BBN bound is combined with the requirement that the resulting signal is ultimately detectable in the audio band.  }
\label{FIGURE6}      
\end{figure}
In the left plot of Fig. \ref{FIGURE6} we illustrated the bound of Eq. (\ref{ABS2}); the shaded area describes the region where 
the bound of Eq. (\ref{ABS2}) is satisfied. It is actually possible to distinguish 
two different shadings. The slightly larger area has been deduced by from the 
numerical result leading to the left plots of Fig. \ref{FIGURE5}. The inner area has been instead 
obtained from the analytic approximation (see also the right plot of  Fig. \ref{FIGURE5}). The comparison of the two results shows that, as long as the ultra high frequency spikes 
are concerned, the approximation developed here is sufficiently accurate 
within an order of magnitude. 

The bound of Eq. (\ref{ABS2}) is actually dominated by the highest frequencies of the spectrum
and if we now move from the THz domain to the audio band the conclusions are quantitatively 
different but qualitatively very similar. For purposes of illustration the current limits on the presence of relic graviton backgrounds in the audio band \cite{LIGO0,LIGO0a,LIGO0b,LIGO1} (see also \cite{MG1}) can be combined with a minimal requirement on the detectability of relic gravitons; we can then require
\begin{equation}
10^{-13} \leq h_{0}^2 \, \Omega_{gw}(\nu_{LVK}, \tau_{0}) < 10^{-10}, \qquad\qquad \nu_{LVK} \leq {\mathcal O}(100)\,\, \mathrm{Hz},
\label{BB3}
\end{equation}
where $\nu_{LVK}$ denotes the Ligo-Virgo-Kagra frequency. The factor $10^{-13}$ is purely illustrative  since we cannot foresee when 
the corresponding sensitivity will be reached by wide-band detectors. In the right plot of Fig. \ref{FIGURE6} the condition of Eq. (\ref{BB3}) has been illustrated with the same logic 
pursued in the derivation of the left plot of the same figure. The two areas are practically 
superimposed; the rightmost region is obtained from the analytic approximation while 
the leftmost area follows from the numerical results. 
\begin{figure}[!ht]
\centering
\includegraphics[height=7.4cm]{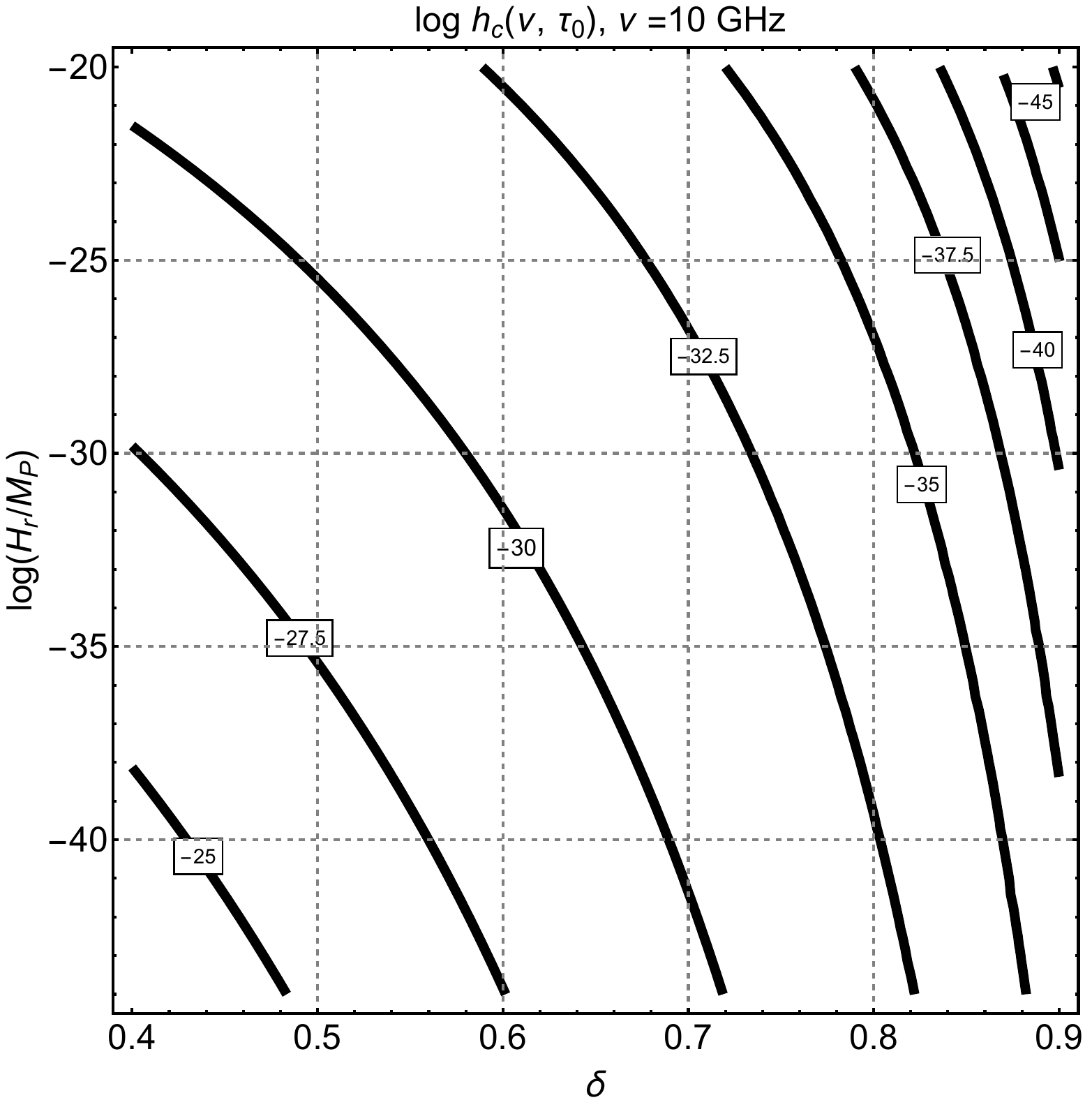}
\includegraphics[height=7.4cm]{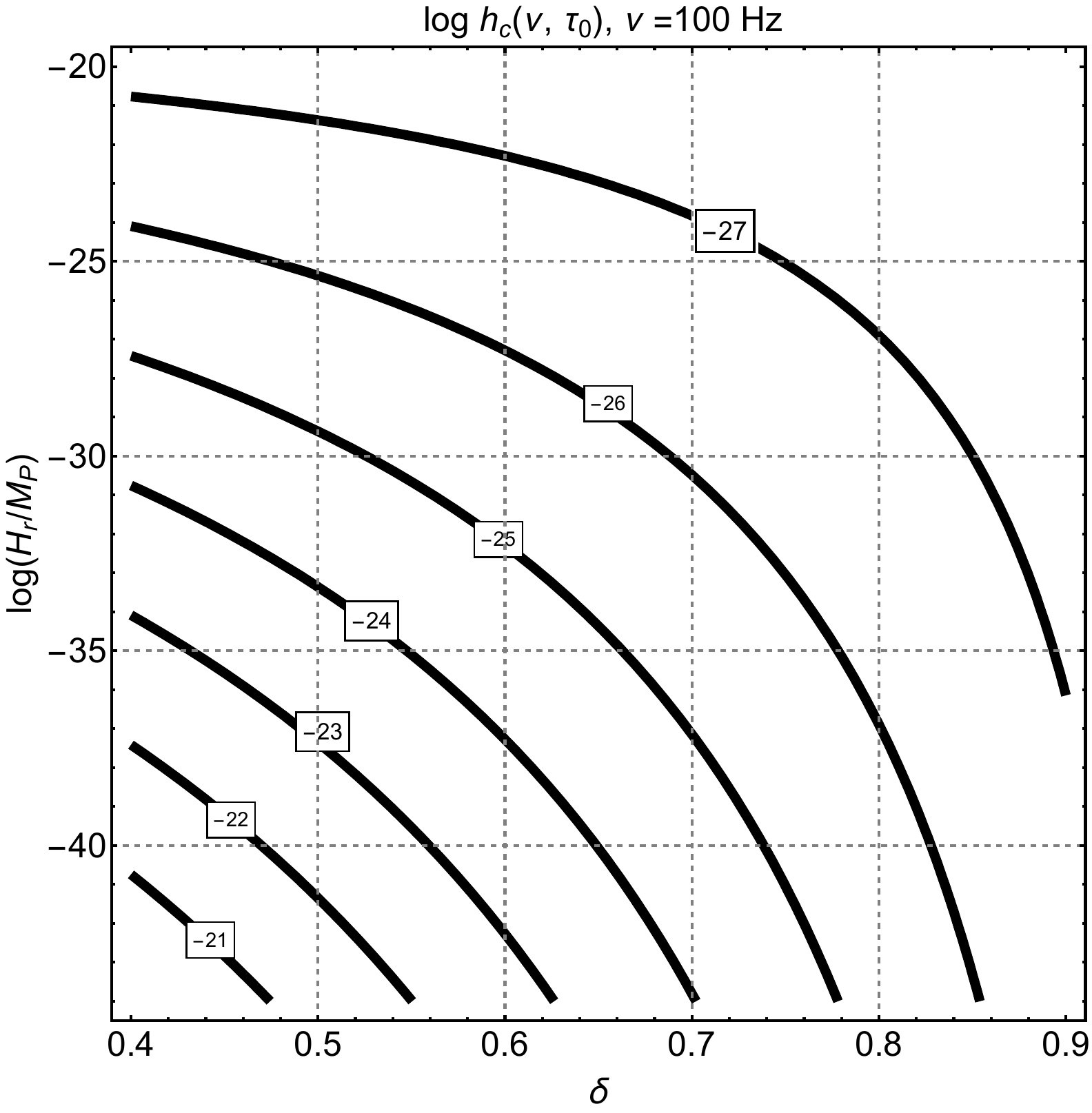}
\caption[a]{The values of the chirp amplitudes 
in the plane $(H_{r}/M_{P}, \, \delta)$ are illustrated for two typical (comoving) frequencies. 
The  labels appearing on the 
contours denote the common logarithms of the chirp amplitudes at the frequencies of $10$ GHz 
(left plot) and of $100$ Hz (right plot). As $\nu$ increases the minimal chirp amplitudes 
potentially measured by a hypothetical high frequency instrument must be reduced in comparison 
with the $h_{c}^{(min)}(\nu,\tau_{0})$ that can be currently measured in audio band. Depending 
on the value of the comoving frequency the minimal chirp amplitude should be $12$ of even $14$ orders 
of magnitude smaller than the ones now associated with a signal in the region of $0.1$ kHz.}
\label{FIGURE7}      
\end{figure}
\subsection{Minimal detectable amplitudes and maximal frequencies}
The analytic and numerical considerations developed in this paper
allow for a direct estimate of the minimal chirp amplitudes required in the ultra-high frequency range for 
a direct detection of the cosmic gravitons. The minimal 
 $h_{c}(\nu,\tau_{0})$ resolved by a hypothetical detector should be at least twelve orders of magnitude smaller than the ones currently measured in the audio band, i.e. in the window ranging between few Hz and the kHz where wide-band detectors are currently operating. To clarify this statement the results obtained so far will now be used by  normalizing the potential signal directly 
in the maximal frequency domain of the spikes. It is important to appreciate that the connection between the spectral energy density (in critical units) and the chirp amplitude $h_{c}(k, \tau_{0})$ is well defined only 
when the relevant wavelengths are shorter than the Hubble radius and this is because 
the chirp amplitude is proportional to the tensor power spectrum $P_{T}(k,\tau)$. Indeed the chirp amplitude 
is simply defined as:
\begin{equation}
\langle \widehat{h}_{i\, j}(\vec{x}, \tau) \, \widehat{h}^{i\, j}(\vec{x}, \tau)\rangle = 2 h_{c}^2(\vec{x},\tau) =
2 \int \frac{d\, k}{k} \,h_{c}^2(k,\tau) j_{0}(k\, r),
\label{HC0a}
\end{equation}
where, as usual, $j_{0}(k\, r)$ is the spherical Bessel function of zeroth order. 
By now recalling Eqs. (\ref{TWOA1}) and (\ref{TWOA4}) we obtain from Eq. (\ref{HC0a}) 
\begin{equation}
\langle \widehat{h}_{i\,j}(\vec{k}, \tau) \, \widehat{h}_{m\,n}(\vec{p}, \tau) \rangle = \frac{4 \,\pi^2}{k^3} \,h_{c}^2(k,\tau) \,{\mathcal S}_{i\,j\,m\,n}(\hat{k})\, \delta^{(3)}(\vec{k}+ \vec{p}),
\label{HC0b}
\end{equation}
which also means that $P_{T}(k,\tau) = 2 \, h_{c}^2(k,\tau)$. Equations (\ref{HC0a})--(\ref{HC0b}) 
implicitly suggest that the only way to obtain a connection between $h_{c}^2(k,\tau)$ 
and the spectral energy density of Eq. (\ref{TWOA9}) is consider, as anticipated, typical
 wavelengths shorter than the Hubble radius; in this case we obtain
 \begin{equation}
 \Omega_{gw}(k,\tau) = \frac{k^2}{6\, H^2\, a^2} \,h_{c}^2(k,\tau), \qquad\qquad Q_{T}(k,\tau) \simeq k^2 \, P_{T}(k,\tau),
 \label{HC0d}
\end{equation}
 where the second (approximate) equation is only valid when the corresponding 
 wavelengths are shorter than the Hubble radius. If we now switch from the comoving wavenumber 
 to the comoving frequency we obtain from Eq. (\ref{HC0d}), at the present time,
 \begin{equation}
\Omega_{gw}(\nu,\, \tau_{0}) = \frac{2\, \pi^2}{3 } \, \biggl(\frac{\nu}{H_{0}}\biggr)^2\,  h_{c}^2(\nu,\tau_{0}) = 1.27 \, \biggl(\frac{h_{0}}{0.7}\biggr)^{-2} \, \biggl(\frac{\nu}{\mathrm{aHz}}\biggr)^{2}\, h_{c}^2(\nu, \tau_{0}).
\label{HC0e}
\end{equation}
Equation (\ref{HC0e}) already explains, in a rather qualitative manner, why the chirp amplitude 
associated with cosmic gravitons should be rather minute in the ultra-high frequency range.

From Eq. (\ref{HC0d}) it is useful to deduce a direct relation between the chirp amplitude 
and the averaged multiplicity in the ultra-high frequency domain:
\begin{equation}
h_{c}(\nu,\tau_{0}) = 7.643 \times 10^{-34}\, \biggl(\frac{\nu}{\mathrm{GHz}}\biggr) \, \sqrt{\overline{n}(\nu, \tau_{0})}.
\label{HC1}
\end{equation}
Since, by definition, around the maximal frequency 
the averaged multiplicity is ${\mathcal O}(1)$, 
the largest spectral domain is only be accessible when $h_{c}^{(min)}(\nu,\tau_{0})$ is smaller (or even much smaller) than ${\mathcal O}(10^{-22})$. We also know that the averaged multiplicity scales as $(\nu/\nu_{max})^{- 4 + n_{T}}$; from 
Eq. (\ref{HC1}) it follows 
\begin{equation}
h^{(min)}_{c}(\nu, \tau_{0}) < 8.13 \times 10^{-32} \biggl(\frac{\nu}{0.1\, \mathrm{THz}}\biggr)^{-1 +n_{T}/2}.
\label{HC2}
\end{equation}
In Fig. \ref{FIGURE7} the labels on the contours correspond to the common logarithm 
of the chirp amplitude in the plane $(H_{r}/M_{P}, \, \delta)$. In the left plot the typical frequency 
has been chosen to be $10$ GHz while in the right plot the frequency falls in the audio band.
From the comparison between the two plots we can argue that the 
the minimal detectable $h_{c}(\nu,\tau_{0})$ at high and intermediate frequencies: to require the same $h_{c}^{(min)}(\nu,\tau_{0})$ both in the audio and in the THz bands makes no sense.
The same argument could be phrased also in terms of the spectral 
energy density since we know that 
\begin{equation}
h_{c}(\nu,\tau_{0}) = 1.263 \times 10^{-20} \biggl(\frac{100 \, \mathrm{Hz}}{\nu}\biggr) \sqrt{h_{0}^2\,
\Omega_{gw}(\nu,\tau_{0})}.
\label{HC3}
\end{equation}
From Eq. (\ref{HC3}) the chirp amplitude can be directly deduced from the analytic determinations 
of $\Omega_{gw}(\nu,\tau_{0})$ and this is what has been done in Fig. \ref{FIGURE7}. 

All in all, a sensitivity ${\mathcal O}(10^{-20})$ or even ${\mathcal O}(10^{-24})$ in the chirp amplitude between the MHz and the THz regions is immaterial for ultra-high frequency gravitons. It has been suggested in the past that microwave cavities \cite{CAV1,CAV2,CAV3,CAV4,CAV5,CAV6} operating in the MHz and GHz regions should be employed for the detection of relic gravitons \cite{EXP1} (see also \cite{CAV6}). The 
relic signal is rather unique for its deep theoretical implications that are probably more relevant that 
the discovery of a new source of gravitational radiation. These  instruments have been also suggested in a series of interesting studies in Refs. \cite{FA1,FA2,FA3,FA4} by setting the required chirp amplitudes in the range $h_{c}^{(min)}(\nu,\tau_{0}) = {\mathcal O}(10^{-20})$ for arbitrarily high frequencies. As suggested in Fig. \ref{FIGURE7} a  minimal chirp amplitude $h^{(min)}_{c}(\nu,\tau_{0}) = {\mathcal O}(10^{-20})$ is more than $10$ orders magnitude larger than the requirements associated with the direct detection of cosmic gravitons. Equations (\ref{HC2})--(\ref{HC3} also clarify why $h_{c}^{(min)}(\nu,\tau_{0})$ must be at least ${\mathcal O}(10^{-32})$ (or smaller) for a potential detection of cosmic gravitons in the THz domain.  We should however 
note that while for $n_{T} > 2$ the largest signal occurs at the largest frequency, for $n_{T} \leq 2$ the signal may increase for frequencies smaller than the THz are . If we consider, for instance, the case $n_{T} \to 1$ we would have that the chirp amplitude in the MHz range could be ${\mathcal O}(10^{-28})$ (this is why this case was regarded as particularly interesting in Refs. \cite{CAV5,CAV6}). Furthermore, when $n_{T} \to 2$ we would have instead that $h_{c}(\nu,\tau_{0})$ is the same at higher and smaller frequencies. Finally for $n_{T} \to 3$  the chirp amplitude at lower frequencies is even more suppressed. There is a non-trivial interplay between the optimal frequency, the features of the signal and the 
noises (especially the thermal one) indicating that the ultra-high frequency 
(close to $\nu_{max}$) is not always the most convenient. This aspect should probably be 
taken into account if the goal is really an accurate assessment of the required sensitivities associated 
with promising (but still hypothetical) high frequency instruments.

\newpage
\renewcommand{\theequation}{6.\arabic{equation}}
\setcounter{equation}{0}
\section{Concluding remarks}
\label{sec6}
The reliable evaluations of the potential signals associated with the ultra-high frequency gravitons must encompass the whole spectrum ranging between few aHz and the THz region. On the one hand the current limits and the known damping sources in the aHz and nHz domains are essential for the analysis of the concordance paradigm and of its neighbouring extensions. On the other hand the production of few pairs of gravitons with opposite three-momenta defines operationally the maximal spectral frequency falling in the THz range. In this broad interval consisting of more than three decades in frequency,  the spectral energy density and the averaged multiplicities of the relic gravitons may offer, in the years to come, a prima facie evidence of the early expansion history of the Universe. Both the presence and the absence of cosmic gravitons domain shall be extremely relevant provided these phenomena are investigated with the appropriate sensitivities: artificially reducing the demands on the 
minimal sensitivities, especially at high frequencies, does not usually offer any effective
shortcut to swift discoveries.

Although below the nHz various sources of suppression concurrently reduce the spectral energy density, we found useful to scrutinize the approximate estimates of the high frequency spikes that may potentially develop between the THz and ${\mathcal O}(300) \sqrt{H_{r}/H_{1}} \, \mathrm{MHz}$ where $H_{r} \geq 10^{-44}\, M_{P}$ (as implied by the tests on the expansion history at the onset of big bang nucleosynthesis) and $H_{1}= {\mathcal O}(10^{-6})\, M_{P}$.  Along this perspective a number of complementary computational schemes have been introduced with the purpose of assessing their mutual accuracy in comparison with more faithful numerical analyses also accounting for the low frequency effects.  We first examined the WKB approach together with various strategies based on the analysis of the transition matrix that relates the evolution of the mode functions between the inflationary and the post-inflationary stages of evolution. Even though between $H_{1}$ and $H_{r}$ various intermediate epochs may appear, in this investigation we simply considered the presence of a single post-inflationary stage preceding the conventional radiation-dominated phase. Exactly the same procedure illustrated here can be however extended to more general situations where the post-inflationary evolution includes different expanding stages. It turns out that the general bounds on the post-inflationary expansion rate can be swiftly deduced with the analytic approach developed in this investigation and they compare quite well with more accurate strategies including the sources of late-time suppression of the spectrum.

For a direct detection of a cosmic signal in the ultra-high frequency band, the present estimates  confirm that the required sensitivity in the chirp amplitude should be at least twelve or even thirteen orders of magnitude smaller than the ones currently accessible between few Hz and the kHz where the wide band interferometers are currently taking data. If electromechanical detectors (e.g. microwave cavities) operating between the MHz and the THz will ever be able to achieve these sensitivities, the timeline of the post-inflationary expansion rate might be observationally accessible in the years to come. As repeatedly suggested in the past, this perspective encourages a synergic approach between experiments scrutinizing different branches of the graviton spectrum at low, intermediate and ultra-high frequencies. It would actually seem rather bizarre to set the detectability strategies for ultra-high frequencies by deliberately disregarding the motivated physical properties of the largest cosmic signal between the MHz and the THz.

\section*{Acknowledgements}
It is also a pleasure to thank A. Gentil-Beccot, A. Kohls, L. Pieper, S. Rohr and J. Vigen of the CERN Scientific Information Service for their kind help.

\end{document}